\documentstyle[11pt]{article}
\def\PRL #1 #2 #3{{\sl Phys.  Rev.  Lett.} {\bf#1} (#2) #3}
\def\NPB #1 #2 #3{{\sl Nucl.  Phys.} {\bf B #1} (#2) #3}
\def\NPBFS #1 #2 #3 #4{{\sl Nucl.  Phys.} {\bf B #2} [FS#1] (#3) #4}
\def\CMP #1 #2 #3{{\sl Commun.  Math.  Phys.} {\bf #1} (#2) #3}
\def\PRD #1 #2 #3{{\sl Phys.  Rev.} {\bf D #1} (#2) #3}
\def\PLA #1 #2 #3{{\sl Phys.  Lett.} {\bf #1A} (#2) #3}
\def\PLB #1 #2 #3{{\sl Phys.  Lett.} {\bf B #1} (#2) #3}
\def\JMP #1 #2 #3{{\sl J.  Math.  Phys.} {\bf #1} (#2) #3}
\def\PTP #1 #2 #3{{\sl Prog.  Theor.  Phys.} {\bf #1} (#2) #3}
\def\SPTP #1 #2 #3{{\sl Suppl.  Prog.  Theor.  Phys.} {\bf #1} (#2) #3}
\def\AoP #1 #2 #3{{\sl Ann.  of Phys.} {\bf #1} (#2) #3}
\def\PNAS #1 #2 #3{{\sl Proc.  Natl.  Acad.  Sci.  USA} {\bf #1} (#2) #3}
\def\RMP #1 #2 #3{{\sl Rev.  Mod.  Phys.} {\bf #1} (#2) #3}
\def\PR #1 #2 #3{{\sl Phys.  Reports} {\bf #1} (#2) #3}
\def\AoM #1 #2 #3{{\sl Ann.  of Math.} {\bf #1} (#2) #3}
\def\UMN #1 #2 #3{{\sl Usp.  Mat.  Nauk} {\bf #1} (#2) #3}
\def\FAP #1 #2 #3{{\sl Funkt.  Anal.  Prilozheniya} {\bf #1} (#2) #3}
\def\FAaIA #1 #2 #3{{\sl Functional Analysis and Its Application}
{\bf #1} (#2) #3}
\def\BAMS #1 #2 #3{{\sl Bull.  Am.  Math.  Soc.} {\bf #1} (#2) #3} 
\def\TAMS #1 #2 #3{{\sl Trans.  Am.  Math.  Soc.} {\bf #1} (#2) #3}
\def\InvM #1 #2 #3{{\sl Invent.  Math.} {\bf #1} (#2) #3}
\def\LMP #1 #2 #3{{\sl Letters in Math.  Phys.} {\bf #1} (#2) #3}
\def\IJMPA #1 #2 #3{{\sl Int.  J.  Mod.  Phys.} {\bf A #1} (#2) #3}
\def\AdM #1 #2 #3{{\sl Advances in Math.} {\bf #1} (#2) #3}
\def\RMaP #1 #2 #3{{\sl Reports on Math.  Phys.} {\bf #1} (#2) #3}
\def\IJM #1 #2 #3{{\sl Ill.  J.  Math.} {\bf #1} (#2) #3}
\def\APP #1 #2 #3{{\sl Acta Phys.  Polon.} {\bf #1} (#2) #3}
\def\TMP #1 #2 #3{{\sl Theor.  Mat.  Phys.} {\bf #1} (#2) #3}
\def\JPA #1 #2 #3{{\sl J.  Physics} {\bf A#1} (#2) #3}
\def\JSM #1 #2 #3{{\sl J.  Soviet Math.} {\bf #1} (#2) #3}
\def\MPLA #1 #2 #3{{\sl Mod.  Phys.  Lett.} {\bf A #1} (#2) #3}
\def\JETP #1 #2 #3{{\sl Sov.  Phys.  JETP} {\bf #1} (#2) #3}
\def\JETPL #1 #2 #3{{\sl Sov.  Phys.  JETP Lett.} {\bf #1} (#2) #3}
\def\PHSA #1 #2 #3{{\sl Physica} {\bf A #1} (#2) #3}
\def\CQG #1 #2 #3{{\sl Class.  Quantum Grav.} {\bf #1} (#2) #3}
\def\SJNP #1 #2 #3{{\sl Sov. J.  Nucl. Phys. (Yadern.Fiz.)} {\bf #1} (#2) #3}
\def\a{\alpha}\def\b{\beta}\def\g{\gamma}\def\d{\delta}\def\e{\epsilon}

\def\k{\kappa}\def\s{\sigma}\def\S{\Sigma}
\def\Th{\Theta}\def\om{\omega}\def\Om{\Omega}\def\G{\Gamma}

\newcommand{\nn}{\nonumber\\}
\newcommand{\p}[1]{(\ref{#1})}
\newcommand{\plabel}{\label}
\begin{document}
\renewcommand{\thefootnote}{\fnsymbol{footnote}}
\setcounter{page}0
\thispagestyle{empty}
\begin{flushright}
{\bf hep-th/9906041 \\
TUW/99-11  \\
1999, 
June 4}\\
\end{flushright}

\begin{center}
{\LARGE 
Superstring 'ending' on super-D9-brane:}
 \\ {\Large 
a supersymmetric action functional 
for the coupled brane system. 
}

\vspace{1.0cm}
\renewcommand{\thefootnote}{\dagger} \vspace{0.2cm}

\bigskip 

{\bf Igor  Bandos}\footnote{Lise Meitner fellow. Also at the 
{\it   Institute for Theoretical Physics, 
NSC Kharkov Institute of Physics and Technology, 
 310108, Kharkov,  Ukraine 
 e-mail: bandos@kipt.kharkov.ua}} 
 and  {\bf Wolfgang Kummer} 

\bigskip 

{\it  Institut f\"{u}r Theoretische Physik, 
\\ 
Technische
 Universit\"{a}t Wien, 
\\ Wiedner Hauptstrasse 8-10, A-1040 Wien\\
 e-mail: wkummer@tph.tuwien.ac.at \\
bandos@tph32.tuwien.ac.at \\
}

\vspace{0.5cm}

{\bf Abstract}
\end{center}

\medskip

{\small

A supersymmetric action functional describing the interaction  of 
the fundamental 
superstring 
with the $D=10$, type $IIB$  Dirichlet super-9-brane is 
presented. 
A set of supersymmetric equations for the coupled system 
is obtained from the action principle. 
It is found that the interaction of the string endpoints with 
the super--D9--brane gauge field requires some restrictions 
for the image of  the gauge field strength. When those restrictions 
are not imposed, the equations imply the absence of the endpoints, 
and the equations coincide either with the ones of the free super-D9-brane
or with the ones for the free closed type $IIB$ superstring.    
Different phases of the coupled system are described.  
A generalization to 
an arbitrary system of intersecting 
branes is discussed. }

\medskip 

\setcounter{page}1
\renewcommand{\thefootnote}{\arabic{footnote}} \setcounter{footnote}0

\newpage

\def\theequation{\thesection.\arabic{equation}}
\section{Introduction}

Intersecting branes and branes ending on branes 
receive much attention now 
\cite{inters1}--\cite{West}
in relation with the development  of  M-theory 
\cite{DbraneM} and 
its application to gauge theories 
\cite{witten96,witten97}. 
However, the studies of \cite{inters1}--\cite{West} were performed 
for the pure bosonic limit of the 
brane systems or for a supersymmetric description in the framework of 
the 'probe brane' approach only. 

In the first case they are based upon the observation  that the ground state 
should not include the nontrivial expectation values of the fermions 
in order to keep (part of) the   Lorentz invariance 
(corresponding to the configuration of the branes).  
Then it is possible to justify that some of the pure bosonic 
solutions  preserve  part  of the target space supersymmetry 
and, just due to this property, saturate the Bogomolnyi bound and 
are thus stable 
(see e.g. \cite{SUGRA}  and refs. therein). 

In the second case one of the branes is treated 
in an the 'external field' of the other brane. The latter can be  
considered either as the solution of low energy supergravity theory 
\cite{inters1,Sato}, or,  in the frame of 
the superembedding approach \cite{bpstv,bsv,hs1,hs2,5-equiv},  
as a superspace, 
where the ends of the probe brane are living 
\cite{Sezgin25,hsc,Sezgin98}. 
In such an approach the $\kappa$-symmetry of the probe brane 
plays an essential role for studying the 'host' brane and the coupled system. 

Despite many successes of these approaches, it is desirable to obtain a 
 complete and manifestly
  supersymmetric description of interacting brane systems.

Of course, the preservation of supersymmetry  
in the presence of boundaries (including the boundaries of open branes 
ending on other branes)  
requires the analysis of anomalies \cite{HW,Mourad}, while at the  
 classical level  the boundary 
breaks  at least half of the supersymmetry \cite{gsw,typeI,Mourad}. 
So at that level
one may search for 
an action for a coupled brane system, which includes manifestly supersymmetric 
bulk terms for all the branes and allows direct variations.
The term 'supersymmetric' will be used below for an action of this type.

The main problem to be faced in a search for such an action is that 
the coordinate functions of intersecting branes 
(or of the open brane and host brane), which define  embeddings  
of their worldvolumes, say 
$${\cal M}^{1+p}= (\zeta ^m ), ~~~ m=0,\ldots p  \qquad 
and \qquad {\cal M}^{1+p'}=(\xi^{m^\prime}), ~~~ m^\prime=0,\ldots p^\prime $$
 into the tangent 
superspace 
$
{\cal M}^{(D~|~N^.2^{[D/2]})}= (X^{\underline{m}}, 
\Theta^{\underline{\mu}I}), ~~~~( \underline{m}= 0, \ldots D-1, ~~~
\underline{\mu}= 1, \ldots 2^{[D/2]}$, 
\\ $\quad I=1, \ldots N)    
$: 
$$ 
{\cal M}^{1+p} \in {\cal M}^{(D~|~N^.2^{[D/2]})} : \qquad 
X^{\underline{m}}=\tilde{X}^{\underline{m}} (\zeta), \qquad 
\Theta^{\underline{\mu}I} = \tilde{\Theta }^{\underline{\mu}I} (\zeta)
$$ 
and 
$$
{\cal M}^{1+p'} \in {\cal M}^{(1+(D-1)|N^.2^{[D/2]})} : \qquad 
X^{\underline{m}}=\hat{X}^{\underline{m}}(\xi ), \qquad 
\Theta^{\underline{\mu}I} = \hat{\Theta}^{\underline{\mu}I} (\xi )
$$ 
should be identified at the intersection 
${\cal M}^\cap \equiv {\cal M}^{1+p} \cap {\cal M}^{1+p'} = 
( \tau ^r ), ~~ r=0, \ldots , dim({\cal M}^\cap )-1 $:
\begin{equation}\plabel{idn}
{\cal M}^\cap \equiv {\cal M}^{1+p} \cap {\cal M}^{1+p'} \in 
{\cal M}^{(D~|~N^.2^{[D/2]})} : \quad  
\tilde{X}^{\underline{m}} (\zeta (\tau ))
=\hat{X}^{\underline{m}} (\xi (\tau ) ), \quad 
\tilde{\Theta }^{\underline{\mu}I} (\zeta (\tau ))= 
\hat{\Theta}^{\underline{\mu}I} (\xi (\tau ) ).  
\end{equation}
Hence the variations $\d \tilde{X} (\zeta), \d \tilde{\Theta } (\zeta )$ 
and $\d \hat{X}(\xi ), \d \hat{\Theta} (\xi )$ may not be considered 
as completely independent.

Recently  we proposed two procedures to solve this problem and   
to obtain a supersymmetric  
action for an interacting  brane system \cite{BK99}. 
One of them provides  the necessary identification \p{idn} by 
the Lagrange multiplier method   
(SSPE or superspace embedded action 
\cite{BK99}). 
Another ('(D-1)-brane dominance' approach or 
Goldstone fermion embedded (GFE) action) 
involves  
a (dynamical or auxiliary) space-time 
filling brane 
and uses the identification of 
all the Grassmann coordinate fields  of lower dimensional branes 
$\hat{\Theta} (\xi )$,  $\tilde{\Theta } (\zeta)$ with the 
images of the (D-1)-brane Grassmann coordinate fields 
\begin{equation}\plabel{idThI}
\hat{\Theta} (\xi ) = \Theta (x(\xi)), \qquad  
\tilde{\Theta} (\zeta ) = \Theta (x(\zeta)).
\end{equation} 
We considered the general properties of the equations of motion 
which follows from such actions  
using the  example of a superstring 
ending on a super--D3--brane. It was  found  that 
both approaches are equivalent and thus justify one the other. 
The super--9--brane was considered as an auxiliary object in \cite{BK99}.
The study of supersymmetric equations of motion for this system will be 
the subject of forthcoming paper \cite{BK99-3}. 

Here we elaborate another example of 
the dynamical system consisting of the fundamental superstring ending on the 
super-D9--brane. We present explicitly the action for 
the coupled system and obtain equations of motion by its direct variation. 

As the super-D9--brane is the space time--filling brane of 
the type $IIB$ superspace, the GFE approach is most natural 
in this case.  Moreover, the system involving the 
dynamical space time--filling brane has some peculiar properties 
which are worth studying 
(see e.g. \cite{Kth}). On the other hand, it 
can be regarded as a relatively simple counterpart of the supersymmetric 
dynamical 
system including superbranes and supergravity 
(see \cite{BK99} for some preliminary considerations).

Several problems arise when one tries   
to find the action for a coupled system of 
the space--time filling superbrane and another super--p--brane. 
The main one is how 
to formulate the supersymmetric generalization of the 
 current (or, more precisely,  
 dual current form)  distributions $J_{D-(p+1)}$
with 
support localized on the brane worldvolume ${\cal M}^{1+p}$.   
Such distributions can be used to present the action of 
a lower dimensional brane as an integral over  the D--dimensional space--time,
 or, equivalently, to the (D-1)--brane worldvolume. 
Then the action for the coupled system  of the lower dimensional branes 
and the space-time filling brane acquires the form of 
an integral over the $D$--dimensional manifold and permits direct 
variation.  

The solution of this problem was presented  in \cite{BK99} and will be 
elaborated here in detail.
For the space-time filling brane the world volume spans the whole bosonic part
 of the target superspace. 
As a consequence, it produces a nonlinear realization of 
the target space supersymmetry. 
The expression for 
the supersymmetry transformations of the bosonic current form 
distributions (which was 
used in \cite{bbs} 
for the description of the interacting bosonic M--branes 
\cite{BSTann,blnpst,schw5}) 
vanishes when the above mentioned identification 
of the Grassmann coordinates of the lower dimensional brane 
with the image of the Grassmann coordinate field of the 
space-time filling brane is imposed. 
This observation provides us with the necessary current distribution form 
and is the key point of our construction\footnote{
It is convenient to first adapt the 
description of the currents to the language of dual current forms, 
whose usefulness had been pointed out  in 
\cite{DL92,Julia}.}.

The second problem  is related to 
the fact that the distributions $J_{D-p-1}$ 
can be used to lift the $(p+1)$ dimensional 
integral to the $D$--dimensional one, 
\begin{equation}\plabel{lift}
\int_{{\cal M}^{1+p} } \hat{{\cal L}}_{p+1} = 
\int_{{\cal M}^{D} } 
J_{D-p-1} \wedge {{\cal L}}_{p+1},  
\end{equation}
only when the integrated $(p+1)$-form $\hat{{\cal L}}_{p+1}$ can be considered 
as a pull--back of a D-dimensional $(p+1)$-form ${{\cal L}}_{p+1}$
living on ${\cal M}^{D}$
onto the $(p+1)$--dimensional surface ${\cal M}^{1+p}\in {\cal M}^{D}$.

Thus, e.g. the superstring Wess-Zumino form 
can be easily 'lifted' up to 
(i.e. rewritten as) the integral over the whole D9-brane worldvolume. 
However, not the entire superstring actions are written as an integral of a 
pull-back of a 10-dimensional form. 
For example, the kinetic term of the Polyakov formulation of 
the (super)string action 
\begin{equation}\plabel{superc}
\int_{{\cal M}^{1+1}}  {\cal L}^{Polyakov}_2  
= \int d^2 \xi \sqrt{-g} g^{\mu\nu} 
\hat{\Pi}_\mu^{\underline{m}}
\hat{\Pi}_\nu^{\underline{n}} \eta_{\underline{m}\underline{n}} 
\equiv 
\int_{{\cal M}^{1+1}} \hat{\Pi}^{\underline{m}}
\wedge * \Pi^{\underline{n}} \eta_{\underline{m}\underline{n}} 
\end{equation}
with 
$$
\hat{\Pi}^{\underline{m}}
\equiv d\hat{X}^{\underline{m}}(\xi)- id\hat{\Th}^I\G^{\underline{m}}
\hat{\Th}^I = d\xi^\mu 
\hat{\Pi}_\mu^{\underline{m}}, \qquad \mu=1,2 \qquad \xi^\mu=(\tau,\s ), 
$$
$$
\eta_{\underline{m}\underline{n}}=diag (+1,-1,\ldots ,-1)
$$
does  not possess such a formulation. Moreover,  it is unclear how to  
define a straightforward  extension of the 2--form ${\cal L}^{Polyakov}_2$
to the whole 10-dimensional space-time. 

The same problem exists for the Dirac-Nambu-Goto and Dirac-Born-Infeld 
kinetic terms of super-Dp-branes and usual super-p-branes.

Though it is possible to treat the 'lifting' relation \p{lift} 
 formally 
(see e.g. \cite{bbs} for a description of bosonic M-branes),  
to address the delicate problems of the supersymmetric coupled 
brane system   
it is desirable to have a version of the superstring and superbrane actions 
which admits a straightforward and explicit lifting  to the whole 
10--dimensional space 
or to the whole D9-brane world volume. 
Fortunately, such a formulation does exist. 
It is the so-called Lorentz harmonic formulation of the superstring 
\cite{BZ} 
 which includes  auxiliary moving frame 
(Lorentz harmonic) variables, treated as  worldsheet fields. 
This is a  geometric  
 (in a sense the first-order) 
action  which can be written in terms 
of differential 
forms without use of the Hodge operation \cite{bpstv,bsv}. 
The only world volume field which is 
not an image of a target space one 
is just the moving frame field (harmonics). 
However, it is possible to extend this field to 
an auxiliary 10-dimensional 
$SO(1,9)/(SO(1,1) \times SO(8)$  'sigma model', which is subject to 
the only condition that it should coincide 
with the  'stringy' harmonics when restricted to the string worldsheet
\footnote{Just the existence of the Lorentz harmonic 
 actions for super--D--branes \cite{bst,baku2,abkz} and super--M--branes 
\cite{bsv,bpst} 
guarantees the correctness of the formal approach 
to the action functional description of interacting bosonic systems   
\cite{bbs}.}.

In this way we obtain a supersymmetric action 
for the interacting system including 
super-D9-brane and a fundamental superstring 'ending' on the 
D9-brane, derive 
the supersymmetric equations of motion directly from the  variation 
of the action and study different phases of the coupled dynamical 
system. 
 We shortly discuss as well the generalization of 
our approach for the case of an arbitrary system of intersecting branes.

For simplicity we are working in  flat target $D=10$ $type ~IIB$ 
superspace. The generalization to brane systems in arbitrary 
supergravity {\sl background} is straightforward. 
 Moreover, our approach allows to involve supergravity in  
an interacting brane system. 
To this end one can include a counterpart of the group manifold action for 
supergravity in the functional describing interacting branes 
instead of (or together with) 
the space--time filling brane  action.
 

In Section 2 we consider the peculiar features of an interacting system 
which contains a space-time filling brane. 
We describe an induced embedding of the 
superstring worldsheet into the D9-brane worldvolume. 
The geometric  action 
\cite{abkz} and the geometric ('first order') form of the supersymmetric 
equations of motion 
for the super--D9--brane are presented in Section 3. 
Section 4 is devoted to the description of the geometric (twistor--like 
Lorentz harmonic) 
action and of the equations of motion for the free type $IIB$ superstring. 
Here Lorentz harmonic variables are used and the issue of supersymmetry 
breaking by boundaries is addressed briefly. 
In Section 5 we introduce the density with support 
localized on the superstring worldsheet and motivate that 
 it becomes invariant under $D=10$ type $IIB$ supersymmetry 
when the identification \p{idThI} is imposed.

The action functional describing the interacting 
system of the super--D9--brane and the (in general open) fundamental  
superstring ('ending' on the super-D9-brane) is presented in Section 6. 
The equations of motion of the interacting  system are derived  
in Section 7  and 
analyzed in 
Section 8.  
The issues of kappa-symmetry and supersymmetry in the coupled system 
are addressed there. 
In the last Section  
we summarize our results 
and also discuss a generalization of our approach to an arbitrary 
system of intersecting branes.

\def\theequation{\thesection.\arabic{equation}}

\section{The space-time filling brane}
\setcounter{equation}{0}

The embedding of the super--D9--brane worldvolume 
\begin{equation}\plabel{D9wv}
{\cal M}^{1+9}= \{ x^m \}, \qquad m=0,\ldots ,9  
\end{equation}
into the $D=10$ type $II$ target superspace 
\begin{equation}\plabel{IIBts}
\underline{{\cal M}}^{(1+9|32)}= \{ X^{\underline{m}},
\Theta^{\underline{\mu}1}, \Theta^{\underline{\mu}2} \}, 
\qquad {\underline{m}}=0,\ldots ,9  
\qquad {\underline{\mu}}=1,\ldots ,16  
\end{equation}
can be described locally by the coordinate superfunctions
\begin{equation}\plabel{superco}
X^{\underline{m}} 
= X^{\underline{m}} (x^m ), \qquad 
\Theta^{I\underline{\mu}} = \Theta^{I\underline{\mu}} (x^m), ~~~ I=1,2 .
\end{equation}
In addition, there is an intrinsic world volume gauge field 
living on the D9-brane world volume
\begin{equation}\plabel{gf}
A = dx^m A_m (x^n) .
\end{equation}

For nonsingular D9-brane configurations 
the function $X^{\underline{m}} (x^m )$ should be assumed 
to be 
nondegenerate in the sense 
$det \left(\partial_n 
X^{\underline{m}} (x^m )\right)\not= 0$. 
Thus the inverse function   
\begin{equation}\plabel{xX}
x^m = x^m (X^{\underline{m}})
\end{equation} 
does exist and, hence, the Grassmann coordinate 
functions \p{superco} and the Born-Infeld gauge field 
\p{gf}
can be considered as functions of 
$X^{\underline{m}}$ variables. 
In this manner an alternative parametrization of the D9-brane world volume is 
provided by
\begin{equation}\plabel{supercVA}
{{\cal M}}^{1+9} ~\rightarrow ~
\underline{{\cal M}}^{(1+9|32)} : \qquad 
{{\cal M}}^{1+9} = 
\{ (X^{\underline{m}}, \Theta^{I\underline{\mu}} (X^{\underline{m}} ) ) \}, 
\qquad A = dX^{\underline{m}} A_{\underline{m}}(X^{\underline{n}}), 
\end{equation}
 which clarifies  the fact that the $D=10$, type $II$ 
super--D9--brane is a theory of Volkov-Akulov Goldstone fermion 
\cite{VA}  combined into a supermultiplet with the  vector field 
$A_{\underline{m}}(X^{\underline{n}})$ (see \cite{abkz}).

Through the intermediate step \p{xX}, \p{supercVA}
we 
can define the {\sl induced embedding of the superstring worldsheet 
into the D9-brane world volume}.

\subsection{
Induced embedding of the superstring  worldsheet}

Indeed, the embedding of the fundamental superstring worldsheet  
\begin{equation}\plabel{IIBwv}
{\cal M}^{1+1}= \{ \xi^{(\pm\pm )} \} 
= \{ \xi^{(++)}, \xi^{(--)} \}, \qquad  \xi^{(++)}= \tau + \s, 
\qquad  \xi^{(--)}= \tau - \s, \qquad  
\end{equation}
into the $D=10$ type $IIB$ target superspace $\underline{{\cal M}}^{(1+9|32)}$
\p{IIBts}
can be described locally by the coordinate superfunctions
\begin{equation}\plabel{supercIIB}
X^{\underline{m}} 
= \hat{X}^{\underline{m}} (\xi^{(\pm\pm )} ), \qquad 
\Theta^{I\underline{\mu}} = \hat{\Theta}^{I\underline{\mu}} (\xi^{(\pm\pm )}), ~~~ I=1,2 .
\end{equation}
However, using the existence of 
the inverse function 
\p{xX}, one can define the {\sl induced} embedding of 
the worldsheet into the D9-brane world volume
\begin{equation}\plabel{x(xi)}
x^m =x^m (\xi )\equiv x^m \left(  \hat{X}^{\underline{m}} (\xi )\right). 
\end{equation}
 
As superstring and super--D9--brane live in the same 
$D=10$ type $IIB$ superspace, we can use the identification 
 of the Grassmann coordinate fields 
of the superstring with the images of the Grassmann coordinate fields 
of the super--D9--brane (Goldstone fermions) on the worldsheet  
\begin{equation}\plabel{X9Xs}
\hat{\Theta}^{I\underline{\mu}} (\xi^{(\pm\pm )})= 
\Theta^{I\underline{\mu}} (\hat{X}^{\underline{m}} (\xi^{(\pm\pm )} )), 
\end{equation}
or, equivalently, 
\begin{equation}\plabel{X9xXs}
\hat{X}^{\underline{m}} (\xi^{(\pm\pm )} )= 
\hat{X}^{\underline{m}} \left(x^m(\xi^{(\pm\pm )} )\right),  
\qquad 
\hat{\Theta}^{I\underline{\mu}} (\xi^{(\pm\pm )})= 
\Theta^{I\underline{\mu}}\left(x^m(\xi^{(\pm\pm )} )\right),  
\end{equation}
to study the interaction of the 
fundamental superstring with the super-D9-brane. 

The approach based on such an identification was called 
'Goldstone fermion embedded' (GFE) in \cite{BK99} 
because, from another viewpoint, the 
superstring worldsheet 
can be regarded as embedded into Goldstone fermion theory 
rather than into superspace.

\subsection{Tangent and cotangent space.}

The pull-backs of the basic forms (flat supervielbeine)
of flat $D=10$ type $IIB$ superspace 
\begin{equation}\plabel{Ea9}
E^{\underline{a}} = \Pi^{\underline{m}}u_{\underline{m}}^{~\underline{a}} 
\equiv  (dX^{\underline{m}}
- i d\Th^1 \s^{\underline{m}} \Th^1 - i d\Th^2 \s^{\underline{m}} \Th^2)
u_{\underline{m}}^{~\underline{a}},  
\end{equation}
 \begin{equation}\plabel{Eal9}
E^{\underline{\a} 1} = d\Theta^{1\underline{\mu}} v_{\underline{\mu}}
^{~\underline{\a}} , 
\qquad 
E^{\underline{\a} 2} = d\Theta^{2\underline{\mu}} v_{\underline{\mu}}
^{~\underline{\a}} 
\end{equation}
 to the D9-brane worldvolume  
are defined by the decomposition  
on the holonomic basis $dx^m$ or $dX^{\underline{m}}$
\begin{equation}\plabel{dXdx}
dX^{\underline{m}}= dx^m \partial_m X^{\underline{m}}(x).
\end{equation}
The basic relations are 
\begin{equation}\plabel{Pipb}
\Pi^{\underline{m}}= 
  dX^{\underline{n}} \Pi^{~\underline{m}}_{\underline{n}}
= dx^m  \Pi^{~\underline{m}}_{{m}}, \qquad 
\end{equation}
\begin{equation}\plabel{Pi1}
\Pi^{~\underline{m}}_{\underline{n}}
\equiv \d^{\underline{m}}_{\underline{n}}
- i \partial_{\underline{n}} 
\Th^1 \s^{\underline{m}} \Th^1 - i \partial_{\underline{n}} 
\Th^2 \s^{\underline{m}} \Th^2 , 
\end{equation}
\begin{equation}\plabel{Pi2}
\Pi^{~\underline{m}}_{{n}}
\equiv \partial_n X^{\underline{m}}
- i \partial_{{n}} 
\Th^1 \s^{\underline{m}} \Th^1 - i \partial_{{n}} 
\Th^2 \s^{\underline{m}} \Th^2 , 
\end{equation}
\begin{equation}\plabel{dThpb}
  d\Theta^{\underline{\mu}I} 
= dX^{~\underline{m}}
\partial_{\underline{m }}\Theta^{\underline{\mu}I} = 
dx^m  \partial_m\Theta^{\underline{\mu}I} (x^n)  , \qquad 
\end{equation}
\begin{equation}\plabel{dThpbd}
  \partial_m\Theta^{\underline{\mu}I} (x^n)  
\equiv \partial_m X^{~\underline{m}} (x)
\partial_{\underline{m }}\Theta^{\underline{\mu}I} . \qquad 
\end{equation}

The matrices $u^{\underline{a}}_{\underline{m}}$, 
$v^{\underline{\a}}_{\underline{\mu}}$, involved into Eqs. 
\p{Ea9}, \p{Eal9} 
take their values in the Lorentz group 
\begin{equation}\plabel{uin}
u^{\underline{a}}_{\underline{m}} \qquad \in SO(1,D-1),  
\end{equation}
and in its doubly covering group $Spin(1,D-1)$
\begin{equation}\plabel{vin}
v^{\underline{\a}}_{\underline{\mu }} \qquad \in Spin(1,D-1),  
\end{equation}
respectively and represent the same Lorentz transformations. The latter 
 imply  the relations 
\begin{equation}\plabel{uvgv}
u^{\underline{a}}_{\underline{m}} 
\tilde{\s}_{\underline{a}}
^{\underline{\a}\underline{\b}} = 
 v^{\underline{\a}}_{\underline{\mu }}
\tilde{\s}_{\underline{a}}
^{\underline{\mu}\underline{\nu}} 
 v^{\underline{\b}}_{\underline{\nu }}, \qquad 
u^{\underline{a}}_{\underline{m}} 
\tilde{\s}^{\underline{m}}
_{\underline{\mu}\underline{\nu}} = 
 v^{\underline{\a}}_{\underline{\mu }}
\tilde{\s}^{\underline{m}}
_{\underline{\a}\underline{\b}} 
 v^{\underline{\b}}_{\underline{\nu }}, \qquad 
\end{equation}
between these matrices 
which reflect the invariance of the $D=10$ sigma matrices 
under the Lorentz group transformations (see \cite{BZ,bpstv}). 
The variables \p{uin}, \p{vin} are not necessary for the 
description of the super--D9--brane itself.
However, as we shall see below, they are useful for the description
of the coupled system including a brane 'ending' on (interacting with) 
the D9-brane.

For that 
system 
it is important to note that the pull-backs of 
bosonic supervielbein forms 
$\Pi^{\underline{m}}$ or $E^{\underline{a}}$ 
of type $IIB$ superspace  
can be used as a basis in the space cotangent to the world-volume of 
D9--brane. In other words, it 
is convenient to use the invertibility of the matrix  
$\Pi_n^{~\underline{m}}$ \p{Pi2} and the harmonic variables 
to define the covariant basis  
$\nabla_{\underline{m}}$  and the one $\nabla_{\underline{a}}$  
of the space tangent to the D9-brane worldvolume by
\begin{equation}\plabel{ddec}
d \equiv dx^n \partial_m = 
dX^{\underline{m}} \partial_{\underline{m}}= 
\Pi^{\underline{m}} \nabla_{\underline{m}} = 
E^{\underline{a}} \nabla_{\underline{a}},  
\end{equation}
\begin{equation}\plabel{der}
\nabla_{\underline{m}} = 
\Pi^{-1}{}^{~\underline{n}}_{\underline{m}} 
\partial_{\underline{n}}= 
\Pi^{-1}{}^n_{~\underline{m}}\partial_n, \qquad  
\nabla_{\underline{a}} \equiv 
u_{\underline{a}}^{~\underline{m}}\nabla_{\underline{m}}. 
\end{equation}

\bigskip

\section{Geometric action 
and equations of motion for super-D9-brane}
 \setcounter{equation}{0}

\subsection{Geometric action}

 The 
geometric action  
for the super--D9--brane 
in flat $D=10$, type $IIB$ superspace 
is \cite{abkz}
\begin{equation}\plabel{SLLL} 
S = \int_{{\cal{M}}^{10}} {\cal{L}}_{10} = 
 \int_{{\cal{M}}^{10}} ({\cal{L}}^0_{10} +  {\cal{L}}^1_{10} +  
{\cal{L}}^{WZ}_{10})  
\end{equation}
where 
\begin{equation}\plabel{L0}
{\cal{L}}^0_{10} =  
\Pi^{\wedge 10} |\eta + F |
\end{equation}
with
$$
|\eta + F | \equiv 
\sqrt{-det(\eta_{\underline{m}\underline{n}}+F_{\underline{m}\underline{n}})}, 
$$
\begin{equation}\plabel{Pi10}
 \Pi^{\wedge 10} \equiv {1 \over (10)!} \e_{\underline{m}_1\ldots 
\underline{m}_{10}} 
\Pi^{\underline{m}_1} \wedge ... \wedge \Pi^{\underline{m}_{10}}, 
\end{equation}

\begin{equation}\plabel{L1}
{\cal{L}}^1_{10} = Q_{8} \wedge (dA-B_2  - {1 \over 2} \Pi^{\underline{m}} 
\wedge 
\Pi^{\underline{n}} F_{\underline{n}\underline{m}}). 
\end{equation}
$A= dx^m A_m(x)$ is the gauge field inherent to the Dirichlet branes, 
$B_2$ represents the NS--NS gauge field with  flat superspace value 
\begin{equation}\plabel{B2def}
B_2 = i \Pi^{\underline{m}}\wedge 
\left(
d\Th^1\s_{\underline{m}}\Th^1  - d\Th^2\s_{\underline{m}}\Th^2 
\right) + 
d\Th^1\s^{\underline{m}}\Th^1 \wedge  d\Th^2\s_{\underline{m}}\Th^2 
\end{equation}
and  field strength 
\begin{equation}\plabel{H3def}
H_3= dB_2 = i \Pi^{\underline{m}}\wedge 
\left(
d\Th^1\s_{\underline{m}} \wedge d\Th^1  - d\Th^2\s_{\underline{m}}\wedge 
d\Th^2 \right). 
\end{equation}
The Wess-Zumino Lagrangian form is the same as 
the one appearing in the standard formulation 
\cite{c1,schw,c2,schw1,bt} 
\begin{equation}\plabel{LWZ}
{\cal{L}}^{WZ}_{10} = e^{{\cal F}_2} \wedge  C \vert_{10} , 
\qquad C = \oplus _{n=0}^{5} C_{2n} , \qquad 
e^{{\cal F}_2}= \oplus _{n=0}^{5} {1\over n!} {\cal F}_2^{\wedge n}, 
\end{equation}
where the formal sum of the RR superforms $C= C_0 + C_2 +...$ 
and of the external powers of the 2--form 
\begin{equation}\plabel{calFD9}
{\cal F}_2 \equiv dA - B_2
 \end{equation} 
(e.g. ${\cal F}^{\wedge 2}\equiv {\cal{F}} \wedge {\cal{F}}$ etc.) 
is used and $\vert_{10}$ means the restriction to the $10$--superform input. 
Let us note that the restriction of the same expression 
\p{LWZ} to the $(p+1)$--form input (where $p=2k-1$ is odd) 
\begin{equation}\plabel{LWZp+1}
{\cal{L}}^{WZ}_{p+1} = e^{\cal{F}} \wedge  C \vert_{p+1}  
= \oplus _{n=0}^{5} C_{2n} \wedge \oplus _{n=0}^{5} {1\over n!} 
{\cal F}^{\wedge n} \vert_{p+1} 
\end{equation}
describes the Wess-Zumino term of the super-Dp-brane of type $IIB$ theory 
\cite{c1,c2,bt}. 
This will be important for the description of the supersymmetric 
generalization of the Born-Infeld equations for the D9-brane gauge fields, 
where the D7-brane Wess-Zumino term appears.

For most applications only the external derivative of the Wess-Zumino term 
is important. It has the form  
\begin{equation}\plabel{dLWZ}
d{\cal{L}}^{WZ}_{10} 
= e^{\cal{F}} \wedge  R \vert_{11} , 
\qquad R = \oplus _{n=0}^{5} R_{2n+1} , \qquad 
\end{equation}
with the 'vacuum' (i.e. flat target superspace) values 
of the Ramond-Ramond curvatures specified as 
\begin{equation}\plabel{RRR}
R = \oplus _{n=0}^{5} R_{2n+1} =  
 e^{- {\cal F}} \wedge   d( e^{\cal{F}} \wedge  C) 
= 2i d\Theta^{2\underline{\nu} }  \wedge  d\Theta^{1\underline{\mu} } \wedge 
\oplus _{n=0}^{4} \hat{\s}^{(2n+1)}_{\underline{\nu}\underline{\mu} }. \qquad 
\end{equation}

\bigskip 

In the action variations and expressions for currents 
the  notion of 'dual' forms  
\begin{eqnarray}\plabel{dualf}
&& \Pi^{\wedge 9}_{\underline{m}} \equiv 
{1 \over 9!} \e_{\underline{m}\underline{m}_1\ldots\underline{m}_{9}} 
\Pi^{\underline{m}_1} \wedge ... \wedge \Pi^{\underline{m}_{9}}, 
 \nonumber \\ 
&& \Pi^{\wedge 8}_{\underline{m}\underline{n}} \equiv 
{1 \over 2 ^. 8!} \e_{\underline{m}\underline{n}\underline{m}_1\ldots 
\underline{m}_{8}} 
\Pi^{\underline{m}_1} \wedge ... \wedge \Pi^{\underline{m}_{8}}, 
\qquad 
\ldots
 \\ 
&& \Pi^{\wedge (10-k)}_{\underline{m}_1\ldots \underline{m}_k} \equiv 
{1 \over k!(10-k)! } \e_{\underline{m}_1\ldots \underline{m}_k
\underline{n}_1\ldots\underline{n}_{(10-k)}} 
\Pi^{\underline{n}_1} \wedge ... \wedge \Pi^{\underline{n}_{(10-k)}} 
\qquad 
 \nonumber 
\end{eqnarray}
is useful. 
The  list of products of the forms \p{dualf} includes the 
useful identities 
\begin{eqnarray}\plabel{dualft}
\Pi^{\wedge 9}_{\underline{m}}  \wedge \Pi^{\underline{n}}  = 
- \Pi^{\wedge 10} \d_{\underline{m}}^{\underline{n}}, \qquad 
\Pi^{\wedge 8}_{\underline{m}\underline{n} } \wedge \Pi^{\underline{l}} = 
\Pi^{\wedge 9}_{[\underline{m} } \d_{\underline{n}]}^{\underline{l}}, \qquad 
 \nonumber 
\\ 
\Pi^{\wedge 7}_{\underline{m}\underline{n} \underline{k} } 
\wedge \Pi^{\underline{l}} = - 
\Pi^{\wedge 8}_{[\underline{m}\underline{n}  } 
\d_{\underline{k}]}^{\underline{l}},
\Pi^{\wedge 6}_{\underline{m}\underline{n} \underline{k} \underline{l} } 
\wedge \Pi^{\underline{r}} =  
\Pi^{\wedge 7}_{[\underline{m}\underline{n} \underline{k}  } 
\d_{\underline{l}]}^{\underline{r}}, \qquad  \\ 
\Pi^{\wedge 6}_{\underline{m}\underline{n} \underline{k} \underline{l} } 
\wedge \Pi^{\underline{r}} \wedge \Pi^{\underline{s}} =  
\Pi^{\wedge 8}_{[\underline{m}\underline{n}  } 
\d_{\underline{k}}^{\underline{r}} \d_{\underline{l}]}^{\underline{s}}.  
\qquad \nonumber 
\end{eqnarray}

\subsection{Variation of geometrical action for D9-brane}

The simplest way to vary the geometrical 
action \p{SLLL}--\p{L1} starts by taking 
the external derivative of the Lagrangian form 
${\cal L}_{10}$ (cf. \cite{bst,abkz})
\begin{equation}\plabel{dL10}
d{\cal L}_{10}= \left(dQ_8 + d{\cal L}_{8}^{WZ}
\vert_{{\cal F}_2\rightarrow F_2 } \right) 
\wedge  \left({\cal F}_2-F_2 \right) + 
 \end{equation}
$$
+ (Q_8 - \Pi^{\wedge 8}_{\underline{n}\underline{m}}
\sqrt{|\eta + F|} (\eta + F)^{-1~\underline{n}\underline{m}}) \wedge 
\Big(- {1 \over 2} \Pi^{\underline{m}}
\wedge \Pi^{\underline{n}} \wedge d
F_{\underline{n}\underline{m}} 
- i \Pi^{\underline{m}} 
\wedge (d\Theta^1 \s^{\underline{n}} \wedge d\Theta^1) 
(\eta - F)_{\underline{n}\underline{m}} +
$$
$$
+ i \Pi^{\underline{m}} 
\wedge (d\Theta^2 \s^{\underline{n}} \wedge d\Theta^2 )
(\eta + F)_{\underline{n}\underline{m}}
\Big)
$$
$$
+ i \Pi^{\wedge 9}_{\underline{m}}
\sqrt{|\eta + F|} (\eta + F)^{-1~\underline{m}\underline{n}} 
\s_{\underline{n}}{}_{\underline{\mu}\underline{\nu}} 
 \wedge  
\left( d\Theta^{2\underline{\mu} }- d\Theta^{1\underline{\rho}}
h_{\underline{\rho}}^{~\underline{\mu} }\right) 
 \wedge  
\left( d\Theta^{2\underline{\nu} }- d\Theta^{1\underline{\s}}
h_{\underline{\s}}^{~\underline{\nu} }\right) + 
$$
$$
+ {\cal O} \left( ({\cal F}_2 - F_2) ^{\wedge 2} )\right), 
$$
where 
${\cal F}_2 \equiv dA- B_2$ \p{calFD9} and  
$F_2 \equiv  {1 \over 2} \Pi^{\underline{m}}
\wedge \Pi^{\underline{n}} 
F_{\underline{n}\underline{m}}$.  Note that 
${\cal F}_2 -F_2$ vanishes due to the algebraic equation 
which is implied by the Lagrange multiplier $Q_8$.
This is the reason why the terms proportional to the second and higher 
(external) powers of $({\cal F}_2 - F_2)$ are indicated by 
${\cal O} \left( ({\cal F}_2 - F_2) ^{\wedge 2} )\right)$
but not written explicitly.

Then we can use the seminal formula 
\begin{equation}\plabel{delatL10}
\d {\cal L}_{10}= i_\d d{\cal L}_{10} + 
d (i_\d {\cal L}_{10})
 \end{equation}
(usually applied for  coordinate variations only) supplemented by 
the formal definition of the contraction with variation symbol
\begin{equation}\plabel{idelatdef}
i_\d d\Theta^{1, 2\underline{\nu} } = 
\d \Theta^{1, 2\underline{\nu} } , \quad 
i_\d \Pi^{\underline{m}} = \d X^{\underline{m}} - 
i \d \Theta^{1}\G^{\underline{m}} \Theta^{1}
- i \d \Theta^{2}\G^{\underline{m}} \Theta^{2}, 
 \end{equation}
\begin{equation}\plabel{idelatdef1}
i_\d dA = \d A , \qquad i_\d dQ_8 = \d Q_8 , \qquad 
i_\d dF_{\underline{m}\underline{n}} = \d F_{\underline{m}\underline{n}} , 
\qquad \ldots . 
 \end{equation}

To simplify the algebraic calculations, one notes that it is 
sufficient to 
write such a formal contraction modulo terms proportional to the square of the 
algebraic equations (the latter remains the same 
for the coupled system as well, because the auxiliary fields, e.g.  $Q_8$,  
do not appear in the action of other branes): 
\begin{eqnarray}\plabel{dS10}
&& \d S_{D9} = \int_{{\cal M}^{1+9}}  
(Q_8 - \Pi^{\wedge 8}_{\underline{k}\underline{l}}
\sqrt{|\eta + F|} (\eta + F)^{-1~\underline{k}\underline{l}}) \wedge 
\Big(- {1 \over 2} \Pi^{\underline{m}}
\wedge \Pi^{\underline{n}} \wedge \d
F_{\underline{n}\underline{m}} + \ldots \Big)+ \nonumber \\ 
&& + \int_{{\cal M}^{1+9}} 
\left(\d Q_8 + ... \right) 
\wedge  \left({\cal F}_2-F_2 \right) +  \nonumber \\ 
&& +\int_{{\cal M}^{1+9}}  \left(d Q_8 + 
d{\cal L}_{8}^{WZ}
\vert_{{\cal F}_2\rightarrow F_2 } \right) 
\wedge  
\left(\d A - i_\d 
B_2 + \Pi^{\underline{n}} F_{\underline{n}\underline{m}} i_\d 
\Pi^{\underline{m}}\right) + \\ 
  && + 2 i \int_{{\cal M}^{1+9}} \Pi^{\wedge 9}_{\underline{m}}
\sqrt{|\eta + F|} (\eta + F)^{-1~\underline{m}\underline{n}} 
\s_{\underline{n}}{}_{\underline{\mu}\underline{\nu}} 
 \wedge  
\left( d\Theta^{2\underline{\mu} }- d\Theta^{1\underline{\rho}}
h_{\underline{\rho}}^{~\underline{\mu} }\right) ~~ 
\left( \d \Theta^{2\underline{\nu} }- \d \Theta^{1\underline{\s}}
h_{\underline{\s}}^{~\underline{\nu} }\right)  
+ \nonumber \\ 
 && + i \int_{{\cal M}^{1+9}} \Pi^{\wedge 8}_{\underline{k}\underline{m}}
\sqrt{|\eta + F|} (\eta + F)^{-1~\underline{k}\underline{n}} 
\s_{\underline{n}}{}_{\underline{\mu}\underline{\nu}} 
 \wedge  
\left( d\Theta^{2\underline{\mu} }- d\Theta^{1\underline{\rho}}
h_{\underline{\rho}}^{~\underline{\mu} }\right) 
 \wedge  
\left( d\Theta^{2\underline{\nu} }- d\Theta^{1\underline{\s}}
h_{\underline{\s}}^{~\underline{\nu} }\right) 
i_\d \Pi^{\underline{m}}  \nonumber  
\end{eqnarray}
Here the terms denoted by $\ldots$  produce contributions to the equations of
motion which are proportional to the algebraic equations and are thus 
inessential.

The spin-tensor matrix $h_{\underline{\mu}}^{~\underline{\nu}}$ entering  
Eqs. \p{dS10} 
is related to the antisymmetric tensor $F_{\underline{n}\underline{m}}$ 
by the Cayley image relations 
\begin{equation}\plabel{hinSpin}
h_{\underline{\mu}}^{~\underline{\nu}} \qquad \in \qquad Spin(1,9) ,
\end{equation}
\begin{equation}\plabel{Id1}
(h\s^{\underline{m}} h^T)_{\underline{\mu}\underline{\nu}} = 
\s^{\underline{n}}_{\underline{\mu}\underline{\nu}} 
k_{\underline{n}}^{~\underline{m}}  \equiv 
 \s^{\underline{n}}_{\underline{\mu}\underline{\nu}} 
(\eta+F)^{-1}_{\underline{n}\underline{l}} 
(\eta-F)^{\underline{l}\underline{m}} , 
\qquad
\end{equation}
\begin{equation}\plabel{Id2}
 k_{\underline{n}}^{~\underline{m}} 
= (\eta+F)^{-1}_{\underline{n}\underline{l}} 
(\eta-F)^{\underline{l}\underline{m}} \equiv 
(\eta-F)_{\underline{n}\underline{l}}(\eta+F)^{-1~\underline{l}\underline{m}} 
\qquad \in \qquad SO(1,9).  
\end{equation}
For more details we refer to \cite{abkz}. 

It is important that $\d A$ enters the compact expression \p{dS10}
for the variation of the super--D9--brane action only 
in  the combination 
\begin{equation}\plabel{susydA}
i_\d \left({\cal F}_2 - F_2 \right) 
\equiv 
\left(\d A - i_\d 
B_2 + \Pi^{\underline{n}} F_{\underline{n}\underline{m}} i_\d 
\Pi^{\underline{m}}\right).  
\end{equation}
It can be called a supersymmetric variation of the gauge field as 
the condition $i_\d \left({\cal F}_2 - F_2 \right)=0$ actually determines 
the supersymmetric transformations of the gauge fields 
(cf. \cite{c1,bst,abkz}). 
Together with 
\p{idelatdef}, the expression \p{susydA} defines the basis of supersymmetric 
variations, whose use simplifies in an essential manner  
the form of the equations of motion. 

 \bigskip 

The formal external derivative 
of the Lagrangian form \p{dL10} can be used as well 
for the general coordinate variation of the action 
\p{SLLL}-\p{L1} 

\begin{equation}\plabel{dS10c}
\d S_{D9}= \int_{{\cal M}^{1+9}} 
\d x^m i_m d{\cal L}_{10}, 
 \end{equation}
where for any q-form $Q_q$ the operation 
$i_m$ is defined by 
\begin{equation}\plabel{imQq}
Q_q= {1 \over q!} dx^{m_1} \wedge \ldots \wedge 
dx^{m_q} Q_{m_q \ldots m_1} \qquad 
i_m Q_q= {1 \over (q-1)!} dx^{m_1} \wedge \ldots \wedge 
dx^{m_{q-1}} Q_{mm_{q-1}\ldots m_1}. \qquad 
 \end{equation}
For the free super--D9--brane such a variation vanishes  identically
when the 'field' equations of motion are taking into account. 
This reflects  
 the evident diffeomorphism invariance of the action \p{SLLL}.  
It is not essential as well in the study of coupled branes in the present 
approach, while in another approach for the description of coupled superbranes 
\cite{BK99} 
such variations play an important role. 

\subsection{Equations of motion for  super--D9--brane}

The equations of motion from the geometric action 
\p{SLLL}--\p{L1} split into the algebraic ones  
obtained from the variation of auxiliary fields
$Q_8$ and $F_{\underline{m}\underline{n}}$
\begin{equation}\plabel{delQ8}
{\cal F}_2 \equiv dA- B_2 = F_2 \equiv  {1 \over 2} \Pi^{\underline{m}}
\wedge \Pi^{\underline{n}} 
F_{\underline{n}\underline{m}}, 
\end{equation}
\begin{equation}\plabel{delF}
Q_8 = \Pi^{\wedge 8}_{\underline{n}\underline{m}}
\sqrt{|\eta + F|} (\eta + F)^{-1~\underline{n}\underline{m}}
\end{equation}
and 
the dynamical ones 
\begin{equation}\plabel{delA}
dQ_8 + d{\cal L}_8^{WZ-D7} = 0 ,
\end{equation}

\begin{equation}\plabel{dTh2eq}
\Pi^{\wedge 9}_{\underline{m}}
\sqrt{|\eta + F|} (\eta + F)^{-1~\underline{m}\underline{n}} 
\s_{\underline{n}}{}_{\underline{\mu}\underline{\nu}} 
 \wedge  
\left( d\Theta^{2\underline{\nu} }- d\Theta^{1\underline{\rho}}
h_{\underline{\rho}}^{~\underline{\nu} }\right) = 0 . 
\end{equation}

If one takes into  account the expression for 
$Q_8$ \p{delF}, the identification 
of $F$ with the gauge field strength \p{delQ8}, as well as the expression 
for the D7-brane Wess-Zumino term 
\begin{equation}\plabel{WZD70}
d{\cal L}_8^{WZ-D7} = 
 e^{\cal{F}} \wedge  R \vert_{9} , 
\qquad R = \oplus _{n=0}^{5} R_{2n+1} 
= 2i d\Theta^{2\underline{\nu} }  \wedge  d\Theta^{1\underline{\mu} } \wedge 
\oplus _{n=0}^{4} \hat{\s}^{(2n+1)}_{\underline{\nu}\underline{\mu} }, \qquad 
\end{equation}
one finds that \p{delA} is just the supersymmetrized Born-Infeld equation.

The fermionic equations \p{dTh2eq} appear as a result of the 
variation with respect to 
$\Theta^2$, while 
the variation with respect to 
$\Th^1$ does not produce any independent equations. 
This fact reflects the Noether identity corresponding to 
the local fermionic $\kappa$--symmetry 
 of the super--D9--brane action \p{SLLL} 
\cite{abkz}.

The explicit {\sl irreducible} form of the D9-brane $\kappa$-symmetry 
transformation can be written with the help of 
the spin--tensor field $h$ \p{hinSpin} -- \p{Id2} \cite{abkz} as  

\begin{equation}\plabel{kD9}
\d\Theta^{1\underline{\mu} }  =\k^{\underline{\mu} }, \qquad 
\d\Theta^{2\underline{\mu} }  = \k^{\underline{\nu} } 
h_{\underline{\nu} }^{~\underline{\mu} }  
\end{equation}
$$
i_\d \Pi^{\underline{m} } = 0 , \qquad \Leftrightarrow \qquad 
\d X^{\underline{m} } = 
i\d\Theta^{1}
\s^{\underline{m} } \Theta^{1} - 
i\d\Theta^{2}
\s^{\underline{m} } \Theta^{2}, \qquad  
$$
\begin{eqnarray}\plabel{dkA}
&& i_\d {\cal{F} } = 0   ~~ \Leftrightarrow \qquad 
\nonumber \\
&& \d A =  i_\d B_2 \equiv i \Pi^{\underline{m}}\wedge 
\left(
\d\Th^1\s_{\underline{m}} \wedge \Th^1  - \d\Th^2\s_{\underline{m}}\wedge 
\Th^2 \right) + \\ 
&& + d\Th^1\s^{\underline{m}}\Th^1 \wedge  \d\Th^2\s_{\underline{m}}\Th^2 
- \d\Th^1\s^{\underline{m}}\Th^1 \wedge  d\Th^2\s_{\underline{m}}\Th^2 , 
\nonumber 
\end{eqnarray}
$$
\d F_{\underline{m}\underline{n}} = 
2i (\eta - F)_{\underline{l}[\underline{m}}
\left(
\nabla_{\underline{n}]}  \Th^1~\s^{\underline{l}}\d \Th^1  
- 
\nabla_{\underline{n}]}\Th^2~\s^{\underline{l}}\d \Th^2 \right), \qquad 
\d Q_8 = 0. 
$$

\bigskip 

The Noether identity reflecting the evident diffeomorphism invariance of the 
action \p{SLLL} is the dependence of the 
equations obtained by varying the action \p{SLLL} with respect to 
 $X^{\underline{m}}(x)$ 
\begin{equation}\plabel{dX}
i \Pi^{\wedge 8}_{\underline{n}\underline{m}}
\sqrt{|\eta + F|} (\eta + F)^{-1~\underline{n}\underline{k}} 
\s_{\underline{k}}{}_{\underline{\mu}\underline{\nu}} 
 \wedge  
\left( d\Theta^{2\underline{\nu} }- d\Theta^{1\underline{\rho}}
h_{\underline{\rho}}^{~\underline{\nu} }\right) 
 \wedge  
\left( d\Theta^{2\underline{\nu} }- d\Theta^{1\underline{\s}}
h_{\underline{\s}}^{~\underline{\nu} }\right) = 0. 
\end{equation}
Indeed it can be proved that Eq. \p{dX} is satisfied identically, when 
Eq.\p{dTh2eq} is taken into account. 

\bigskip

Turning back to the fermionic equations 
\p{dTh2eq}, let us note that after decomposition 
\begin{equation}\plabel{dThdec}
d\Theta^{2\underline{\nu} }- d\Theta^{1\underline{\rho}}
h_{\underline{\rho}}^{~\underline{\nu} }= \Pi^{\underline{m}} 
\Psi_{\underline{m}}^{~\underline{\nu}}, 
\end{equation}
where 
\begin{equation}\plabel{Psidef}
\Psi_{\underline{m}}^{~\underline{\nu}}= 
\nabla_{\underline{m}}
\Theta^{2\underline{\nu} }- \nabla_{\underline{m}}\Theta^{1\underline{\rho}}
h_{\underline{\rho}}^{~\underline{\nu} }
\end{equation}
and $ \nabla_{\underline{m}}$ defined by 
$d= dx^m \partial_m = \Pi^{\underline{m}}\nabla_{\underline{m}}$ 
\p{ddec}, \p{der}, 
one arrives for \p{dTh2eq} at \cite{abkz}
\begin{equation}\plabel{eqPsi}
- i \Pi^{\wedge 10}
\sqrt{|\eta + F|} 
\s_{\underline{k}}{}_{\underline{\mu}\underline{\nu}} 
\Psi_{\underline{m}}^{~\underline{\nu}}
(\eta + F)^{-1~\underline{m}\underline{n}} =0, 
\qquad  
\Leftrightarrow \qquad 
\s_{\underline{k}}{}_{\underline{\mu}\underline{\nu}} 
\Psi_{\underline{m}}^{~\underline{\nu}}
(\eta + F)^{-1~\underline{m}\underline{n}} =0. 
\end{equation}

\bigskip

\section{Geometric action 
and free 
equations of motion for type $IIB$ superstring}
\setcounter{equation}{0}

\subsection{Geometric action and moving frame variables 
(Lorentz harmonics)}

In the geometric action for type IIB superstring  \cite{BZ,bpstv,bsv} 
\begin{equation}\plabel{acIIB} 
S_{IIB}=
\int\limits^{}_{{\cal M}^{(1+1)}} \hat{{\cal L}}_2^{IIB} = 
\int\limits^{}_{{\cal M}^{(1+1)}}
\left({1\over 2} \hat{E}^{++} \wedge \hat{E}^{--}   
- \hat{B}_{2} \right) , 
\end{equation}
the hat (as in \p{acIIB}) 
indicates the fields restricted to 
(or living on) a superstring worldsheet \p{IIBwv}
$
{\cal M} ^{(1+1)}= \{ \xi^{(\pm\pm)} \}~~
$.  
$~\hat{B}_2$ is the pull-back of the NS-NS 
gauge field with the 'vacuum' (i.e. flat superspace) value  
\p{B2def} which  plays the role of the Wess-Zumino term in the superstring 
action, furthermore 
\begin{equation}\plabel{Epm}
\hat{E}^{\pm\pm} = \hat{\Pi}^{\underline{m}} 
\hat{u}_{\underline{m}}^{\pm\pm},  
\qquad     
\end{equation} 
where 
$\hat{u}_{\underline{m}}^{\pm\pm}(\xi )$ are vector harmonics 
\cite{sok,BZ,bpstv,bsv}, i.e. two light--like vector fields  
entering the $SO(1,9)$ valued matrix \p{uin}
\begin{equation}\plabel{harmv}
\hat{u}^{\underline{a}}_{\underline{m}} (\xi ) 
\equiv ( 
\hat{u}^{++}_{\underline{m}}, \hat{u}^{--}_{\underline{m}}, 
\hat{u}^{i}_{\underline{m}}
) \qquad \in \qquad SO(1,D-1).  
\end{equation} 
This matrix describes a moving frame attached to the worldsheet and 
thus provides the possibility to adapt the general bosonic vielbein 
of the flat superspace to the embedding of the worldsheet 
\begin{equation}\plabel{Eunda}
\hat{E}^{\underline{a}}= \left(  
\hat{E}^{++}, \hat{E}^{--}, \hat{E}^{i}\right) \equiv 
\hat{\Pi}^{\underline{m}} \hat{u}_{\underline{m}}^{~\underline{a}}. 
\qquad     
\end{equation}

The properties of the harmonics \p{harmv}  
are collected in the  Appendix A. To obtain equations of motion from the 
geometric action \p{acIIB} it is important that the variations of 
the light-like harmonics 
$\hat{u}_{\underline{m}}^{\pm\pm}$
should be performed with the constraint \p{harmv}, i.e. with 
\begin{equation}\plabel{harmvc}
\hat{u}^{\underline{a}}_{\underline{m}} \eta_{\underline{m}\underline{n}}  
\hat{u}^{\underline{b}}_{\underline{n}} =
\eta^{\underline{a}\underline{b}}~~~ \Rightarrow  
~~~\cases{
 u^{++}_{\underline{m}}
u^{++\underline{m}} = 0 , ~~
u^{--}_{\underline{m}} u^{--\underline{m}} =0, \cr  
u^{~i}_{\underline{m}}
u^{++\underline{m}} = 0 , ~~
u^{i}_{\underline{m}}
u^{--\underline{m}} =0 , \cr
u^{++}_{\underline{m}} u^{--\underline{m}} =2 , ~~~
u^{i}_{\underline{m}}
u^{j\underline{m}} = - \d^{ij}
\cr } , 
\end{equation}
taken into account. The simplest way to implement this consists in  solving 
 the conditions of the conservation of the constraints \p{harmvc} 
$$
\d \hat{u}^{~\underline{a}}_{\underline{m}} \eta_{\underline{m}\underline{n}}  
\hat{u}^{~\underline{b}}_{\underline{n}} 
+
\hat{u}^{~\underline{a}}_{\underline{m}} \eta_{\underline{m}\underline{n}}  
\d \hat{u}^{~\underline{b}}_{\underline{n}} = 0
$$
with respect to $\d \hat{u}^{~\underline{b}}_{\underline{n}}$
and, thus, to  
define 
 a set of variations
('admissible variations' \cite{BZ})
\footnote{ This is the place to note that a  
similar technique was used (\cite{deWit} and refs. therein) 
in the study of 
the $G/H$ sigma model fields, appearing in the maximal $D=3,4,5$
supergravities 
($G/H = E_{8(+8)}/SO(16), E_{7(+7)}/SU(8),$ $E_{6(+6)}/USp(8)$).   
}
 which 
then shall be treated as independent. 
Some of those variations 
$i_\d f^{++i}$, $i_\d f^{--i}$, $i_\d \om$
 enter the expression for the admissible variations of the 
light-like harmonics \cite{BZ}  
 \begin{equation}\plabel{du++}
\d {u}^{++}_{\underline{m}} = 
u^{++}_{\underline{m}} i_\d \om 
+ \hat{u}^{i}_{\underline{m}}  i_\d f^{++i}, \qquad 
\d u^{--}_{\underline{m}} = -\hat{u}^{--}_{\underline{m}} i_\d \om 
+ u^{i}_{\underline{m}}  i_\d f^{--i}, \qquad 
\end{equation}  
while the other  $i_\d A^{ij}$ are involved in the variations 
of the orthogonal components of a moving frame 
 \begin{equation}\plabel{dui}
\d u^{i}_{\underline{m}} = - u^{j}_{\underline{m}}  i_\d A^{ji} +
 {1\over 2} u_{\underline{m}}^{++} i_\d f^{--i} +
 {1\over 2} u_{\underline{m}}^{--} i_\d f^{++i} (d)
\end{equation} 
only and thus produce no inputs into the variation of the action 
\p{acIIB}.

The derivatives of the harmonic 
variables should dealt with in the same way.

\subsection{Action variation and equations of motion}

The external derivative of the Lagrangian form ${\cal L}_2$ 
is 
\begin{equation}\plabel{dLIIB} 
d{\cal L}_2^{IIB} = -2i E^{++} \wedge E^{-1}_{\dot{q}}\wedge E^{-1}_{\dot{q}} 
+ 2i E^{--} \wedge E^{+2}_{{q}}\wedge E^{+2}_{{q}} + 
\end{equation}
$$
+ {1 \over 2} E^i \wedge 
\left( E^{--} \wedge f^{++i} - E^{++} \wedge f^{--i} 
+ 4i  ( E^{+1}_{{q}}\wedge E^{-2}_{\dot{q}} - 
E^{+2}_{{q}}\wedge E^{-1}_{\dot{q}}) \g^i_{q\dot{q}}\right). 
$$
Here 
\begin{equation}\plabel{f++i} 
f^{++i} \equiv
u^{++}_{\underline m} d u^{\underline{m}i} , \qquad
f^{--i} \equiv
u^{--}_{\underline m} d u^{\underline{m}i} , \qquad
\end{equation}
\begin{equation}\plabel{omA} 
\om \equiv {1 \over 2}
u^{--}_{\underline m} d u^{\underline{m}++} , \qquad
A^{ij} \equiv
u^{i}_{\underline m} d u^{\underline{m}j} , \qquad
\end{equation}
are Cartan forms \cite{BZ,bpstv} (see Appendix A) and  
\begin{equation}\plabel{Eundal}
\hat{E}^{\underline{\a }I} \equiv
d\hat{\Th}^{\underline{\mu} I}\hat{v}_{\underline{\mu }}^{\underline{\a }}= 
\left(\hat{E}^{I+}_{~{q}},  
 \hat{E}^{I-}_{~\dot{q}} \right) 
\qquad     
\end{equation} 
$$ q=1, \ldots 8, \qquad \dot{q}=1, \ldots 8, \qquad 
$$
are pull-backs of the fermionic supervielbein forms 
which, together with 
\p{Eunda}, form a basis 
of the flat target superspace.   
They involve the spinor harmonics \cite{gds,BZ} 
\begin{equation}\plabel{harms}
\hat{v}_{\underline{\mu }}^{\underline{\a }}= 
\left(\hat{v}^{I+}_{\underline{\mu }q},  
 \hat{v}^{I-}_{\underline{\mu }\dot{q}} \right) 
\qquad \in \qquad     Spin(1,9)
\end{equation} 
which  represent the same Lorentz rotation (relating  
the 'coordinate frame'  
$\Pi^{\underline{m}}, d\Th^{\underline{\mu }I}$ of the 
target superspace with the arbitrary frame  
$E^{\underline{a}}, E^{\underline{\a }I}$) 
as the vector harmonics \p{harmv} and, hence, are connected with them by 
Eqs.  \p{uvgv}. The latter include in particular the relations 
\begin{equation}\plabel{0harms10}
u^{{++}}_{\underline{m}} \s^{\underline{m}}_{\underline{\mu}\underline{\nu}}
= 
2 v^{~+}_{\underline{\mu}q} v^{~+}_{\underline{\mu}q} , \qquad 
u^{{--}}_{\underline{m}} 
\s^{\underline{m}}_{\underline{\mu}\underline{\nu}}=
2 v^{~-}_{\underline{\mu}\dot{q}} v^{~-}_{\underline{\mu}\dot{q}} , \qquad 
u^{{i}}_{\underline{m}} 
\s^{\underline{m}}_{\underline{\mu}\underline{\nu}}=
2 v^{~+}_{\{ \underline{\mu}q} \g^i_{q\dot{q}} 
v^{~-}_{\underline{\nu}\} \dot{q}} , \qquad 
\end{equation}
which were used to write $d{\cal L}_2^{IIB}$ in a compact form 
\p{dLIIB}. 
For further details concerning harmonics we refer 
to Appendix A and to the original references 
\cite{gds,BZ,bpstv}. 

\bigskip  

Now one can 
calculate 
the variation of the action \p{acIIB} of closed type $IIB$ superstring 
from the expression 
\p{dLIIB} using the technique described in Section 2.2 

\begin{eqnarray}\plabel{delacIIB} 
&& \d S_{IIB}=
\int\limits^{}_{{\cal M}^{(1+1)}} i_\d d{\cal L}_2^{IIB} = 
\int\limits^{}_{{\cal M}^{(1+1)}}
{1\over 2} \hat{E}^{i} \wedge 
\left( E^{--} i_\d f^{++i} - E^{++} i_\d f^{--i} + \ldots \right) + 
\nonumber \\ 
&& \int\limits^{}_{{\cal M}^{(1+1)}} 
\left(  \hat{M}_{2}^i ~u_{\underline{m}}^i   
+  2i \hat{E}^{1+}_{~{q}}
\wedge \hat{E}^{1+}_{~{q}} u_{\underline{m}}^{--}  
- 2i \hat{E}^{2-}_{~\dot{q}} \wedge \hat{E}^{2-}_{~\dot{q}}  
u_{\underline{m}}^{++} \right)
i_\d \Pi^{\underline{m}} +
\\ 
&& 
+ \int\limits^{}_{{\cal M}^{(1+1)}} 
\left(-4i \hat{E}^{++} \wedge \hat{E}^{1-}_{~\dot{q}}
v_{\underline{\mu}\dot{q}}^{~-} \d \Theta^{1\underline{\mu}} + 
4i \hat{E}^{--} \wedge \hat{E}^{2+}_{~{q}}
v_{\underline{\mu}{q}}^{~+} \d \Theta^{2\underline{\mu}}\right). 
\nonumber 
\end{eqnarray}
Here 
\begin{equation}\plabel{hM2def}
\hat{M}_{2}^i \equiv 
{1\over 2} \hat{E}^{--} \wedge \hat{f}^{++i} - 
{1\over 2} \hat{E}^{++} \wedge \hat{f}^{--i} 
+  2i \hat{E}^{1+}_{~{q}}\wedge \g^i_{q\dot{q}}\hat{E}^{1-}_{~\dot{q}}
- 2i \hat{E}^{2+}_{~{q}}\wedge \g^i_{q\dot{q}}\hat{E}^{2-}_{~\dot{q}}
\end{equation} 
and the dots in the first line denote the terms 
$$
f^{++i} i_\d E^{--} - f^{--i} i_\d E^{++} +   
4i \left(\hat{E}^{1+}_{~{q}} \g^i_{q\dot{q}} 
\hat{v}_{\underline{\mu}\dot{q}}^{~-} 
+  \hat{E}^{1-}_{~\dot{q}} \g^i_{q\dot{q}} 
\hat{v}_{\underline{\mu}{q}}^{~+} \right) \d \Theta^{1\underline{\mu}} 
- 4i \left(\hat{E}^{2+}_{~{q}} \g^i_{q\dot{q}} 
\hat{v}_{\underline{\mu}\dot{q}}^{~-} 
+  \hat{E}^{2-}_{~\dot{q}} \g^i_{q\dot{q}} 
\hat{v}_{\underline{\mu}{q}}^{~+} \right) \d \Theta^{2\underline{\mu}}
$$ 
which produce contributions proportional to $E^i$ into the action variation. 
They are essential only when we search for 
the $\kappa$--symmetry of the free type $IIB$ superstring action. 

It is worth mentioning that, in contrast to the standard formulation 
\cite{gsw}, 
the geometric action \p{acIIB} possesses the {\sl irreducible} 
$\kappa$--symmetry 
whose transformation is given by (cf. \cite{BZ,bsv})
\begin{equation}\plabel{kappastr} 
\d \hat{\Th}^{\underline{\mu}1} = \k^{+q}  
\hat{v}^{-\underline{\mu}}_{\dot{q}}, \qquad 
\d \hat{\Th}^{\underline{\mu}2} = \k^{-\dot{q}}
\hat{v}^{+\underline{\mu}}_{{q}}, 
\end{equation}
$$
\d \hat{X}^{\underline{m}} 
= i \d\hat{\Th}^{\underline{\mu}1}
\s^{\underline{m}}\hat{\Th}^{\underline{\mu}1} +
 i \d\hat{\Th}^{\underline{\mu}2}
\s^{\underline{m}}\hat{\Th}^{\underline{\mu}2}
$$
\begin{equation}\plabel{kappav}
\d \hat{v}_{\underline{\mu}q}^{~+} = {1 \over 2} i_\d f^{++i} 
\g^i_{q\dot{q}} \hat{v}_{\underline{\mu}\dot{q}}^{~-}, \qquad 
\d \hat{v}_{\underline{\mu}\dot{q}}^{~-} = {1 \over 2} i_\d f^{--i} 
\g^i_{q\dot{q}} \hat{v}_{\underline{\mu}q}^{~+}  \qquad 
\end{equation}
$$
\d \hat{u}_{\underline{m}}^{++} = \hat{u}_{\underline{m}}^{i}  i_\d f^{++i} , 
\qquad \d \hat{u}_{\underline{m}}^{--} = \hat{u}_{\underline{m}}^{i}  
i_\d f^{--i} , 
\qquad \d \hat{u}_{\underline{m}}^{i} = 
{1 \over 2} \hat{u}_{\underline{m}}^{++}  i_\d f^{--i} + 
{1 \over 2} \hat{u}_{\underline{m}}^{--}  i_\d f^{++i},  
$$
(cf. Appendix A) with  $i_\d f^{++i},  i_\d f^{--i}$ determined by 
\begin{eqnarray}\plabel{kappaf}
&& \hat{E}^{++} i_\d f^{--i} - \hat{E}^{--} i_\d f^{++i} = \\ 
&& = 
4i \left(\hat{E}^{1+}_{~{q}} \g^i_{q\dot{q}} 
\hat{v}_{\underline{\mu}\dot{q}}^{~-} 
+  \hat{E}^{1-}_{~\dot{q}} \g^i_{q\dot{q}} 
\hat{v}_{\underline{\mu}{q}}^{~+} \right) \d \Theta^{1\underline{\mu}} 
- 4i \left(\hat{E}^{2+}_{~{q}} \g^i_{q\dot{q}} 
\hat{v}_{\underline{\mu}\dot{q}}^{~-} 
+  \hat{E}^{2-}_{~\dot{q}} \g^i_{q\dot{q}} 
\hat{v}_{\underline{\mu}{q}}^{~+} \right) \d \Theta^{2\underline{\mu}}. 
\nonumber 
\end{eqnarray}

The equations of motion for the free closed 
type $IIB$ superstring can be extracted easily from \p{delacIIB}
\begin{equation}\plabel{Ei=0}
\hat{E}^{i} \equiv  \hat{\Pi}^{\underline{m}} 
\hat{u}_{\underline{m}}^{i}=0,  
\qquad     
\end{equation} 
\begin{equation}\plabel{M2=0}
\hat{M}_2^{i} = 0,  
\end{equation} 
\begin{equation}\plabel{Th1str}
\hat{E}^{++} \wedge \hat{E}^{1-}_{~\dot{q}} = 0,  
\qquad     
\end{equation} 
\begin{equation}\plabel{Th2str}
\hat{E}^{--} \wedge \hat{E}^{2+}_{~{q}} = 0 , 
\qquad     
\end{equation} 
where  $\hat{M}_2^{i}$, $E^{\pm\pm}$, $ \hat{E}^{2+}_{~{q}}$, 
${E}^{1-}_{~\dot{q}}$ are defined in \p{hM2def}, 
\p{Epm}, \p{Eundal}, respectively.

\bigskip

\subsection{Linearized fermionic equations}

The proof of  equivalence of the Lorentz harmonic formulation \p{acIIB}  
with the standard action of the Green-Schwarz superstring 
has been given 
in \cite{BZ}.
To make this equivalence intuitively evident, let us consider 
the fermionic equations of 
motion 
\p{Th1str}, \p{Th2str} in the linearized approximation,  
fixing a static gauge 

\begin{equation}\plabel{staticg}
\hat{X}^{\pm\pm} \equiv X^{\underline{m}}u_{\underline{m}}^{\pm\pm}= 
\xi^{(\pm\pm )}. 
\end{equation} 
Moreover, we use the $\kappa$--symmetry \p{kappastr} 
to remove half the components 
$
\hat{\Th}^{1+}_{~{q}} = 
\hat{\Th}^{\underline{\mu}1} \hat{v}^{~+}_{\underline{\mu}{q}}, 
~\hat{\Th}^{2-}_{~\dot{q}} = 
\hat{\Th}^{\underline{\mu}2} \hat{v}^{~-}_{\underline{\mu}\dot{q}}, 
$ 
of the Grassmann coordinate fields  
\begin{equation}\plabel{kappag}
\hat{\Th}^{1+}_{~{q}} = 
\hat{\Th}^{\underline{\mu}1} \hat{v}^{~+}_{\underline{\mu}{q}}= 0,  
\qquad \hat{\Th}^{2-}_{~\dot{q}} = 
\hat{\Th}^{\underline{\mu}2} \hat{v}^{~-}_{\underline{\mu}\dot{q}}=0.  
\end{equation} 

Thus we are left with  $8$ bosonic and $16$ fermionic fields   
\begin{equation}\plabel{analb}
\hat{X}^{i} = 
\hat{X}^{\underline{\mu}} \hat{u}^{~i}_{\underline{m}},  \qquad  
\hat{\Th}^{1-}_{~\dot{q}} = 
\hat{\Th}^{\underline{\mu}1} \hat{v}^{~-}_{\underline{\mu}\dot{q}}, 
\qquad 
\hat{\Th}^{2+}_{~{q}} = 
\hat{\Th}^{\underline{\mu}2} \hat{v}^{~+}_{\underline{\mu}{q}}.  
\end{equation}

In the linearized approximation all the inputs from the derivatives of 
harmonic variables (i.e. Cartan forms \p{f++i}, \p{omA}) 
disappear from the fermionic equations for the physical Grassmann 
coordinate fields. 
Thus we arrive at the counterpart of the gauge fixed string 
theory in the light--cone gauge. 
Then it is not hard to see that Eqs. \p{Th1str}, \p{Th2str} reduce 
to the opposite chirality conditions for physical fermionic fields
\begin{equation}\plabel{linfeq}
\partial_{--} \hat{\Th}^{1-}_{~\dot{q}} = 0 , \qquad 
\partial_{++}\hat{\Th}^{2+}_{~{q}} = 0. 
\end{equation} 

To obtain the bosonic equations, the derivatives of the harmonics 
(Cartan forms \p{f++i}) must be taken into account. 
After exclusion of the auxiliary variables one obtains 
that Eqs. \p{Ei=0}, \p{M2=0} reduce to the usual 
free field equations for $8$ bosonic fields $X^i$ 
(see Appendix B for details)
\begin{equation}\plabel{linbeq}
\partial_{--} 
\partial_{++}\hat{X}^{i} = 0. 
\end{equation}

\bigskip

\subsection{Geometric action with boundary term}

To formulate the interaction of the open superstring 
with the super--D9--brane we have to add to the action \p{acIIB} 
the boundary term which describes the coupling to the 
gauge field 
$A=dx^m A_m(x)$ inherent to the D9-brane (see \p{L1}, \p{LWZ}, \p{calFD9}, 
\p{LWZp+1}). 
Thus the  complete action for the open 
fundamental  superstring becomes (cf. \cite{Sezgin98})
\begin{equation}\plabel{acIIBb} 
S_{I}= S_{IIB} + S_{b} = 
\int\limits^{}_{{\cal M}^{(1+1)}}  \hat{{\cal L}}_{2} =
\int\limits^{}_{{\cal M}_{(1+1)}}
\left({1\over 2} \hat{E}^{++} \wedge \hat{E}^{--}   
- \hat{B}_{2} \right) + \int\limits^{}_{\partial {\cal M}^{(1+1)}}  
\hat{A} ~. 
\end{equation}

 The variation of the action \p{acIIBb} differs from 
$\int i_\d d {\cal L}_2^{IIB}$ in \p{delacIIB} 
by boundary contributions. 
In the supersymmetric basis \p{idelatdef}, \p{susydA}
the variation becomes 
\begin{eqnarray}\plabel{varSI}
&& \d S_I = \int_{{\cal M}^{1+1}} i_\d d {\cal L}_2^{IIB} + 
\int_{\partial {\cal M}^{1+1}} i_\d 
\left( {\cal F}_2-F_2\right) + 
\nonumber \\
&& +\int_{\partial {\cal M}^{1+1}}
\left( {1 \over 2} {E}^{++} u_{\underline{m}}^{--}
- {1 \over 2} {E}^{--}  u_{\underline{m}}^{++}   
- \Pi^{\underline{m}} F_{\underline{n}\underline{m}}   
\right) i_\d \Pi^{\underline{m}}.  
\end{eqnarray}
It is worth mentioning that 
no boundary contribution 
with  variation $\d \Theta^I$  appears. 
This does not 
contradict the well--known fact that the presence of a 
worldsheet boundary breaks at least a 
half of the target space $N=2$ supersymmetry. 
Indeed,  for the supersymmetry 
transformations 
\begin{equation}\plabel{susy1}
\d_{susy} X^{\underline{m}} = \Theta^{I} \sigma^{\underline{m}} \e^I , \qquad  
\d_{susy} \Theta^{I\underline{\mu}} =  \e^{I\underline{\mu}} 
\end{equation}
the variation 
$i_{\d } \Pi^{\underline{m}}$ is nonvanishing and reads  
\begin{equation}\plabel{Pisusy}
i_{\d_{susy}} \Pi^{\underline{m}}= 2 \d_{susy} X^{\underline{m}}= 
 2 \Theta^{I} \sigma^{\underline{m}} \e^I.  
\end{equation} 
Imposing the boundary conditions 
$\hat{\Theta}^{1\underline{\mu}}(\xi(\tau ))
 = \hat{\Theta}^{2\underline{\mu}} (\xi(\tau ))$ one arrives 
at the conservation of $N=1$ supersymmetry whose embedding into the 
type $IIB$ supersymmetry group is defined by 
$\e^{\underline{\mu}1}=-  \e^{\underline{\mu}2}$. 
Actually these conditions provide 
$i_\d \hat{\Pi}^{\underline{m}} (\xi(\tau ))=0$ and, as a consequence, 
the vanishing of 
the variation \p{varSI} (remember that the 
supersymmetry transformations of the gauge fields 
are defined by $i_\d ({\cal F}_2 -F_2)=0 $). 

The above consideration in the frame of the Lorentz harmonic approach 
results in the interesting observation that 
the supersymmetry breaking by a boundary is related to the 
'classical reparametrization anomaly': indeed the 
second line of the expression \p{varSI}, 
which produces the nonvanishing variation under $N=2$ supersymmetry 
transformation with \p{Pisusy}, contains only 
 $i_\d \Pi^{\underline{m}}$ , which 
can be regarded as parameters of the reparametrization gauge symmetry 
of the free superstring 
($i_\d \Pi^{\underline{m}} u_{\underline{m}}^{\pm\pm}$) 
and free super--D3--brane ($i_\d \Pi^{\underline{m}}$), respectively.

There exists a straightforward way to keep half of the 
rigid target space supersymmetry of the superstring--super-D9-brane system 
by incorporation of the additional boundary  term 
$\int_{\partial{\cal M}^{1+1}} \phi_{1\underline{\mu}} 
\left(
\hat{\Theta}^{1\underline{\mu}}(\xi(\tau ))
 - \hat{\Theta}^{2\underline{\mu}} (\xi(\tau ))  
 \right)$ with a Grassmann Lagrange multiplier one form 
$\phi_{1\underline{\mu}}$ 
(see Appendix A in \cite{BK99}). 
 However, following \cite{Mourad,Sezgin25,BK99}, we accept 
in our present  paper the 
'soft' breaking of the supersymmetry by boundaries 
at the classical level (see \cite{HW,Mourad} for 
symmetry restoration by anomalies). 
We expect that the BPS states preserving  part of the target space 
supersymmetry 
will appear as particular solutions of the coupled superbrane equations
following from our action.

\bigskip

\section{Current forms and unified description of string and D9-brane }
\setcounter{equation}{0}

\subsection{Supersymmetric current form}

For a simultaneous description of 
super--D9--brane and fundamental superstring, 
we have to define an 8-form distribution $J_8$ with support 
on the string worldsheet. In the pure bosonic case  
(see e.g. \cite{bbs}) 
one requires 
\begin{equation}\plabel{J8def}
\int_{{\cal M}^{1+1}} \hat{{\cal L}}_2 = 
\int_{\underline{{\cal M}}^{1+9}} J_8 \wedge {\cal L}_2 , 
\end{equation}
where 
\begin{equation}\plabel{L2IIB}	
 {\cal L}_2 = {1 \over 2} dX^{\underline{m}} \wedge dX^{\underline{n}}  
{\cal L}_{\underline{n} \underline{m}} ( X^{\underline{l}} )
\end{equation}
is an arbitrary two-form in the $D=10$ dimensional space-time  
$\underline{{\cal M}}^{1+9}$
and 
\begin{equation}\plabel{L2IIBs}
\hat{{\cal L}}_2 = {1 \over 2}
 d\hat{X}^{\underline{m}} (\xi ) \wedge d\hat{X}^{\underline{n}} (\xi )
{\cal L}_{\underline{n} \underline{m}} ( \hat{X}^{\underline{m}}(\xi)) 
\end{equation}
is its pull-back onto the string worldsheet. 

It is not hard to verify that the appropriate 
expression for the current form 
$J_8$ is given by \cite{bbs}
\begin{equation}\plabel{J80}
J_8 = (dX)^{\wedge 8}_{\underline{n} \underline{m}} 
J^{\underline{n} \underline{m}}( X) = 
{ 1 \over 2! 8!} \e_{\underline{m} \underline{n}\underline{n}_1 \ldots 
\underline{n}_8}  d{X}^{\underline{n}_1 } \wedge \ldots \wedge
dX^{\underline{n}_8} \int_{{\cal M}^{1+1}} 
d\hat{X}^{\underline{m}} (\xi ) \wedge d\hat{X}^{\underline{n}} (\xi )
\d^{10} \left( X - \hat{X} (\xi )\right)~.
\end{equation}
Indeed, inserting \p{J80} and \p{L2IIB} into r.h.s. of \p{J8def}, 
using \p{dualft} and changing the order of integrations, 
after performing the integration over $d^{10}X$ one arrives at the 
l.h.s. of \p{J8def}.

\subsection{Superstring boundaries and current (non)conservation}

If the superstring worldsheet is closed 
($\partial {\cal M}^{1+1}= 0$) 
the current $J^{\underline{m}\underline{n}}$ 
is conserved, i.e. $J_8$ is a closed form  
\begin{equation}\plabel{cl}
\partial {\cal M}^{1+1}= 0 \qquad \Rightarrow \qquad 
dJ_8 = 0, \qquad \Leftrightarrow \qquad 
\partial_{\underline{m}} 
J^{\underline{n} \underline{m}} = 0 .
\end{equation}

For the open (super)string this does not hold. 
Indeed, assuming that the 10-dimensional space 
and the D9-brane worldvolume has no boundaries 
$
\partial {\cal M}^{1+9}= 0, 
$
substituting instead of ${\cal L}_2$ a closed two form, say $dA$, 
and using  Stokes' theorem one arrives at 
\begin{equation}\plabel{dJ8def}
\int_{\partial {\cal M}^{1+1}} A = 
\int_{{\cal M}^{1+1}} dA = 
\int_{{{\cal M}}^{1+9}} J_8 \wedge dA = 
\int_{{{\cal M}}^{1+9}} dJ_8 \wedge A .
\end{equation}
Thus the form $dJ_8$ has  support localized at the boundary of the worldsheet 
(i.e. on the worldline of the string endpoints). 

This again can be justified by an explicit calculation with Eqs. 
\p{J80} and \p{dualft}, which results in 
\begin{equation}\plabel{dJ80}
dJ_8 = 
- (dX)^{\wedge 9}_{\underline{n}} 
\partial_{\underline{m}}  
J^{\underline{m} \underline{n}}( X) = 
- (dX)^{\wedge 9}_{\underline{n}} 
j^{\underline{n}}( X) 
\end{equation}
$$
\partial_{\underline{m}}J^{\underline{m}\underline{n}}
( X)= - j^{\underline{n}}( X), \qquad 
$$
with 
\begin{equation}\plabel{j0}
j^{\underline{n}}( X) \equiv 
\int_{\partial{\cal M}^{1+1}} 
d\hat{X}^{\underline{m}} (\tau ) 
\d^{10} \left( X - \hat{X} (\tau )\right), \qquad 
\end{equation}
{ where the proper time $\tau$ parametrizes the boundary of the string 
worldsheet 
$\partial{\cal M}^{1+1}= \{ \tau \}$.

Actually the boundary of superstring(s) shall have (at least) two 
connected components $\partial{\cal M}^{1+1}= \oplus_{j} {\cal M}_j^1$, each 
parametrized by the own proper time $\tau_j$. Then the rigorous 
expression for the boundary current 
\p{dJ80}, \p{j0} is 
$$
j^{\underline{n}}( X) \equiv 
\S_{j}^{}
\int_{{\cal M}^{1}_j} 
d\hat{X}^{\underline{m}} (\tau_j ) 
\d^{10} \left( X - \hat{X} (\tau_j )\right). \qquad 
$$
We, however, will use the  simplified notations \p{j0} in what follows.}

It is useful to define the local density 1--form $j_1$ 
on the worldsheet with  support on the boundary of 
worldsheet 
\begin{equation}\plabel{j10}
j_1 = d\xi^{\mu} \e_{\mu\nu} \int_{\partial {\cal M}^{1+1}} 
d\tilde{\xi}^{\nu} (\tau ) 
\d^{2} \left( \xi - \tilde{\xi} (\tau )\right)~, \qquad 
\end{equation}
which has the properties 
\begin{equation}\plabel{j1def}
\int_{\partial {\cal M}^{1+1}} \hat{A} \equiv 
\int_{{\cal M}^{1+1}} d\hat{A} = \int_{{\cal M}^{1+1}} j_1 \wedge 
\hat{A} = 
\int_{{\cal M}^{1+9}} dJ_8 \wedge \hat{A} 
\end{equation}
for any 1-form 
$$
A= dX^{\underline{m}} A_{\underline{m}} (X),  
$$
e.g., for the D9-brane gauge field \p{gf} considered in the 
special parametrization 
\p{xX}, \p{supercVA}. 

In the sense of the last equality in \p{j1def} one can write a formal 
relation 
\begin{equation}\plabel{dJ8j1}
dJ_8 = J_8 \wedge j_1 
\end{equation}
(which cannot be treated straightforwardly as the form $j_1$ can not be 
regarded as pull-back of a 10-dimensional 1-form). 

\bigskip

\subsection{Variation of current form distributions and supersymmetry }

The variation of the form 
\p{J80} becomes 
\begin{equation}\plabel{deJ80}
\d J_8 = 
3 (dX)^{\wedge 8}_{[\underline{m} \underline{n}} \partial_{\underline{k}]}
\int_{{\cal M}^{1+1}} 
d\hat{X}^{\underline{m}} (\xi ) \wedge d\hat{X}^{\underline{n}} (\xi )
\left( \d X^{\underline{k}}- \d \hat{X}^{\underline{k}} (\xi ) \right)~ 
\d^{10} \left( X - \hat{X} (\xi )\right)~- 
\end{equation}
$$
- 2 (dX)^{\wedge 8}_{\underline{m} \underline{n}}
\int_{\partial {\cal M}^{1+1}} 
d\hat{X}^{\underline{m}} (\tau ) \left( \d X^{\underline{n}}- 
\d \hat{X}^{\underline{n}} (\tau ) \right)~
\d^{10} \left( X - \hat{X} (\tau )\right)~.
$$

Let us turn to the target space supersymmetry transformations 
\p{susy1}.
For the string coordinate fields it has the form 
\begin{equation}\plabel{susyIIB}
\d \hat{X}^{\underline{m}}(\xi ) = \hat{\Theta}^{I} (\xi ) 
\sigma^{\underline{m}} \e^I , 
\qquad  
\d \hat{\Theta}^{I\underline{\mu}} =  \e^{I\underline{\mu}} 
\end{equation}
while for the super-D9-brane it reads  
\begin{equation}\plabel{susyD9}
\d X^{\underline{m}} (x) = \Theta^{I} (x) \sigma^{\underline{m}} \e^I , 
\qquad  
\d \Theta^{I\underline{\mu}} (x) =  \e^{I\underline{\mu}}.  
\end{equation}

In the parametrization \p{supercVA} corresponding to 
the introduction of the inverse function \p{xX} the transformation 
\p{susyD9} coincides with the Goldstone fermion realization 
\begin{equation}\plabel{susyD9X}
\d X^{\underline{m}} = \Theta^{I} (X) \sigma \e^I , \qquad  
\d \Theta^{I\underline{\mu}} (X) \equiv \Theta^{I\underline{\mu}~\prime} 
(X^{\prime} )- \Theta^{I\underline{\mu}} (X)
= \e^{I\underline{\mu}} 
\end{equation}

Thus, if we identify the $X^{\underline{m}}$ entering the 
current density \p{J80} with the bosonic coordinates of superspace, 
parametrizing the super-D9-brane by \p{xX}, we can use \p{deJ80} 
to obtain  
 the supersymmetry transformations of the current density 
\p{J80}
\begin{equation}\plabel{deJ80susy}
\d J_8 = 
3 (dX)^{\wedge 8}_{[\underline{m} \underline{n}} \partial_{\underline{k}]}
\int_{{\cal M}^{1+1}} 
d\hat{X}^{\underline{m}} (\xi ) \wedge d\hat{X}^{\underline{n}} (\xi )
\left( \Theta^{I} (X) 
- \hat{\Theta}^{I} (\xi ) \right)\sigma^{\underline{k}} \e^I ~
\d^{10} \left( X - \hat{X} (\xi )\right)~- 
\end{equation}
$$
- 2 (dX)^{\wedge 8}_{\underline{m} \underline{n}}
\int_{\partial {\cal M}^{1+1}} 
d\hat{X}^{\underline{m}} (\tau ) \left( \Theta^{I} (X) 
- \hat{\Theta}^{I} 
(\tau ) \right)\sigma^{\underline{n}} \e^I ~
\d^{10} \left( X - \hat{X} (\tau )\right)~.
$$

Now it is evident that the current density 
\p{J80} becomes invariant 
under the  supersymmetry transformations 
\p{susyIIB}, \p{susyD9} after the identification
\begin{equation}\plabel{ThhatTh}
\hat{\Theta}^{I} (\xi )= \Theta^{I} \left(\hat{X} (\xi ) \right) 
\end{equation}
of the superstring coordinate fields  $\hat{\Theta}^{I} (\xi )$
with the image  $\Theta^{I} \left(\hat{X} (\xi ) \right)$ 
of the super--D9--brane coordinate field   
$\Theta^{I} (X)$ 
is implied.

\subsection{Manifestly supersymmetric representations for the 
distribution form}

In the presence of the D9-brane whose world volume spans the whole 
$D=10$ dimensional super-time, 
we can rewrite  \p{J80} as

\begin{equation}\plabel{J81}
J_8 = (dx)^{\wedge 8}_{{n} {m}} 
J^{{n}{m}}( x) = 
{ 1 \over 2! 8!} \e_{{n}{m} {n}_1 \ldots {n}_8}  
d{x}^{{n}_1 } \wedge \ldots \wedge
dx^{{n}_8} \int_{{\cal M}^{1+1}} 
d\hat{x}^{{m}} (\xi ) \wedge d\hat{x}^{{n}} (\xi )
\d^{10} \left( x - \hat{x} (\xi )\right) , 
\end{equation}
where the function $\hat{x} (\xi )$ is defined through 
$\hat{X} (\xi )$ with the use of the inverse function \p{xX}, i.e. 
$
\hat{X} (\xi )= X(\hat{x} (\xi ))
$~, cf. \p{x(xi)}. 

Passing from \p{J80} to \p{J81} the identity  
\begin{equation}\plabel{deltapr}
\d^{10} \left( X - \hat{X}(\xi ) \right) \equiv 
\d^{10} \left( X - X(\hat{x}(\xi )) \right) 
= {1 \over det( {\partial X \over  \partial x})} 
\d^{10} \left( x - \hat{x}(\xi ) \right) 
\end{equation}
has to be taken into account.

The consequences of this observation are two-fold: 
\begin{itemize}
\item {\bf i)} 
We can use $J_8$  
to represent an integral over the string worldsheet as an integral over 
the D9-brane worldvolume 
\footnote{Note the difference of the manifolds involved into the 
r.h.s-s of \p{J8def1} and \p{J8def}. This will be important 
for the supersymmetric case. 
}  
\begin{equation}\plabel{J8def1}
\int_{{\cal M}^{1+1}} \hat{{\cal L}}_2 = 
\int_{{{\cal M}}^{1+9}} J_8 \wedge {\cal L}_2 
\end{equation}
for any 2--form 
\begin{equation}\plabel{L2D9}
 {\cal L}_2 = {1 \over 2} dx^{{m}} \wedge dx^{{n}}  
{\cal L}_{{n} {m}} ( x^{{l}} )
\end{equation}
{\sl living on the D9--brane world volume} ${{\cal M}}^{1+9}$,  
e.g. for the field strength ${\cal F}_2 = dA -B_2$
\p{calFD9} of the D9-brane gauge field \p{gf}. 
The pull--back 
\begin{equation}\plabel{L2D9s}
\hat{{\cal L}}_2 = {1 \over 2}
 d\hat{x}^{{m}} (\xi ) \wedge d\hat{x}^{{n}} (\xi )
{\cal L}_{{n}{m}} ( \hat{x}^{{l}}(\xi)) 
\end{equation}
is defined in \p{J8def1} with the use of 
the inverse function \p{xX}. 

\item {\bf ii)} 
As the coordinates $x^n$ are inert under 
the target space supersymmetry \p{susy1}, 
the current density $J_8$ 
is supersymmetric invariant.
{\sl Hence, when   the identification \p{ThhatTh}  
\begin{equation}\plabel{Thid}
\hat{\Theta}^{I} (\xi)= {\Theta}^{I} \left(\hat{x}(\xi)\right)  
\qquad 
\end{equation}
is made, 
it is possible to use  Eqs. \p{J8def}, \p{J8def1}, \p{J81}    
 to lift the complete superstring  action \p{acIIBb} 
to the 10-dimensional integral form.}  
\end{itemize}

The manifestly supersymmetric form of the current density  
appears after passing to the 
supersymmetric  
basis \p{Pipb}, \p{dThpb} of the  space tangential to 
${\cal M}^{1+9}$. With the decomposition \p{Pipb}  
$J_8$ 
becomes 
\begin{equation}\plabel{J8Pi9}
J_8 = 
(\Pi)^{\wedge 8}_{\underline{n} \underline{m}} 
J_{(s)}^{\underline{n} \underline{m}}( X) = 
{ 1 \over 2! 8!} \e_{\underline{n} \underline{m}\underline{n}_1 \ldots 
\underline{n}_8}  \Pi^{\underline{n}_1 } \wedge \ldots \wedge
\Pi^{\underline{n}_8} { 1 \over det(\Pi_{\underline{r}}^{~\underline{s}})} 
\int_{{\cal M}^{1+1}} 
\hat{\Pi}^{\underline{n}} \wedge \hat{\Pi}^{\underline{m}} 
\d^{10} \left( X - \hat{X} (\xi )\right). 
\end{equation}
In Eq. \p{J8Pi9} the only piece where the supersymmetric invariance is not 
manifest 
is $\d^{10} \left( X - \hat{X} (\xi )\right)$. 
However,  in terms of 
D9-brane world volume coordinates  we arrive at 
\begin{equation}\plabel{J8Pi}
J_8= { 1 \over 2! 8!} \e_{\underline{n} \underline{m}\underline{n}_1 \ldots 
\underline{n}_8}  \Pi^{\underline{n}_1 } \wedge \ldots \wedge
\Pi^{\underline{n}_8} { 1 \over det(\Pi_{{r}}^{~\underline{s}})} 
\int_{{\cal M}^{1+1}} 
\hat{\Pi}^{\underline{m}} \wedge \hat{\Pi}^{\underline{n}} 
\d^{10} \left( x - \hat{x} (\xi )\right) ~, 
\end{equation}
where the determinant in the denominator is calculated for the matrix 
$
\Pi_{{n}}^{~\underline{m}} =  
\partial_{{n}} X^{\underline{m}}(x) - i \partial_{{n}} 
\Th^1 \s^{\underline{m}} \Th^1 - i \partial_{{n}} 
\Th^2 \s^{\underline{m}} 
\Th^2. 
$ \p{Pi2}.

\bigskip

The manifestly supersymmetric expression for the exact dual current 9-form
$dJ_8$ \p{dJ80} is provided by 
\begin{equation}\plabel{dJ8Pi}
dJ_8 \equiv 
 (dx)^{\wedge 9}_{{n}} \partial_m J^{{m}{n}}(x) 
= - { 1 \over det(\Pi_{{r}}^{~\underline{s}})} 
(\Pi)^{\wedge 9}_{\underline{m}} 
\int_{\partial {\cal M}^{1+1}} 
\hat{\Pi}^{\underline{m}} 
\d^{10} \left( x - \hat{x} (\tau )\right). 
\end{equation}
\begin{equation}\plabel{j1}
\partial_{{m}}J^{{m}{n}}(x)= - j^{{n}}( x), \qquad 
j^{{n}}(x) \equiv 
\int_{\partial{\cal M}^{1+1}} 
d\hat{x}^{{n}} (\tau ) 
\d^{10} \left( x - \hat{x} (\tau )\right).
\end{equation}

\bigskip

\section{An action for the coupled system} 
\setcounter{equation}{0}

In order to obtain 
a covariant action for the coupled system 
with the current form $J_8$,  
one more step is needed. 
Indeed, our lifting rules \p{J8def}
with the density $J_8$ \p{J80}, \p{J81} are valid for a form 
${\hat{\cal L}}_2$ which is the pull-back of a form 
${{\cal L}}_2$ living either on the whole $D=10$ type $IIB$ superspace 
\p{L2IIB}, or, 
at least, on the whole 10-dimensional worldvolume of the 
super-D9-brane \p{L2D9}.  
Thus imposing the identification \p{Thid} 
 we can straightforwardly rewrite the Wess-Zumino term $\int \hat{B}_2$ 
and the 
boundary term of the superstring action $\int \hat{A}$ 
as  integrals over the 
super--D9--brane world volume 
$\int J_8 \wedge B_2 + \int dJ_8 \wedge A$. 

But the 'kinetic term' of the superstring action 
 \begin{equation}\plabel{kin}
\int_{{\cal M}^{1+1}} {\hat{{\cal L}}}_0 \equiv 
\int_{{\cal M}^{1+1}} {1 \over 2} \hat{E}^{++}\wedge \hat{E}^{--} 
\end{equation}
with $\hat{E}^{++}, \hat{E}^{--}$ defined by Eqs. \p{Epm}
requires an additional consideration regarding the harmonics \p{harmv}, 
which so far were defined only as 
 worldsheet fields. 

In order to represent the kinetic term \p{kin} as an integral over the 
D9-brane world volume too, 
we have to introduce a counterpart of the harmonic fields
\p{harmv}, \p{harms} 
{\sl in the whole 10-dimensional space or in the D9-brane world volume}
\begin{equation}\plabel{harmvD9}
{u}^{\underline{a}}_{\underline{m}} (x) 
\equiv ( 
{u}^{++}_{\underline{m}}(x), {u}^{--}_{\underline{m}}(x), 
{u}^{i}_{\underline{m}}(x)
) \qquad \in \qquad SO(1,9) 
\end{equation}
\begin{equation}\plabel{harmsD9}
{v}_{\underline{\mu }}^{\underline{\a }}= 
\left({v}^{I+}_{\underline{\mu }q},  
 {v}^{I-}_{\underline{\mu }\dot{q}} \right) 
\qquad \in \qquad  Spin(1,9)    
\end{equation} 
(see \p{harmv}--\p{uvgv}).

Such a 'lifting' of the harmonics to the super--D9--brane worldvolume 
creates the fields 
of an auxiliary ten dimensional 
$SO(1,9)/(SO(1,1) \times SO(8))$ 'sigma model'. The only restriction  for 
these new fields is that they should coincide with the  'stringy' harmonics 
on the worldsheet:  

$
{u}^{\underline{a}}_{\underline{m}} \left(x(\xi )\right) = 
\hat{u}^{\underline{a}}_{\underline{m}} (\xi ) ~: 
$
\begin{equation}\plabel{harmvD9IIB}
{u}^{++}_{\underline{m}}
\left(x(\xi )\right) = \hat{u}^{++}_{\underline{m}} (\xi ), 
\qquad 
\hat{u}^{--}_{\underline{m}}
\left(x(\xi )\right) = \hat{u}^{--}_{\underline{m}}(\xi ), 
\qquad  
{u}^{i}_{\underline{m}}
\left(x(\xi )\right) = \hat{u}^{i}_{\underline{m}} (\xi) 
\end{equation}

$
{v}_{\underline{\mu }}^{\underline{\a }} \left(x(\xi )\right) = 
\hat{v}_{\underline{\mu }}^{\underline{\a }} (\xi) ~: 
$
\begin{equation}\plabel{harmsD9IIB}
{v}^{I+}_{\underline{\mu }q}
 \left(x(\xi )\right) = \hat{v}^{I+}_{\underline{\mu }q} (\xi ) , \qquad 
 {v}^{I-}_{\underline{\mu }\dot{q}} 
 \left(x(\xi )\right) =
 \hat{v}^{I-}_{\underline{\mu }\dot{q}} (\xi ) \qquad     
\end{equation} 

\bigskip 

In this manner we 
arrive at the full 
supersymmetric action describing the coupled 
system  of the open fundamental superstring interacting with 
the super--D9--brane (cf. \p{SLLL}--\p{dLWZ}, \p{acIIB}): 

\begin{eqnarray}\plabel{SD9+IIB} 
&& S =  \int_{{\cal{M}}^{10}} \left( {{\cal{L}}_{10}} 
+ J_8 \wedge {{\cal{L}}_{IIB}} + dJ_8 \wedge A
\right) =
\nonumber  \\
&& = \int_{{\cal{M}}^{10}} 
\left[\Pi^{\wedge 10} \sqrt{-det(\eta_{\underline{m}\underline{n}}
+F_{\underline{m}\underline{n}})} + 
Q_{8} \wedge \left( dA-B_2  - {1 \over 2} \Pi^{\underline{m}} \wedge 
\Pi^{\underline{n}} ~F_{\underline{m}\underline{n}}\right)  +  
e^{\cal{F}} \wedge  C~ \vert_{10} ~\right] + 
 \\ 
&& + \int_{{\cal{M}}^{10}} J_8 \wedge \left({1 \over 2} 
{E}^{++} \wedge {E}^{--}   
- B_{2} \right)  +  \int_{{\cal{M}}^{10}} dJ_8 \wedge A 
\nonumber  
\end{eqnarray}

\bigskip

\section{Supersymmetric equations for the coupled system} 
\setcounter{equation}{0}

\subsection{Algebraic  equations}

The Lagrange multiplier $Q_8$ and auxiliary field 
$F_{\underline{m}\underline{n}}$ are not involved into the superstring action 
\p{acIIBb}, 
while the harmonics are absent in the super--D9--brane part 
\p{SLLL}--\p{L1} of the action
\p{SD9+IIB}. Thus we conclude that  the {\sl algebraic equations} 
\p{delQ8}, \p{delF}, \p{Ei=0} 
are the same as in the free models.

\subsubsection{Equations obtained from varying the harmonics }

Indeed, variation with respect to the harmonics (now extended to the 
whole $D=10$ space time or, equivalently, to the super--D9--brane  
world volume \p{harmvD9}) 
produces the equations 

\begin{equation}\plabel{du(c)}
J_8 \wedge E^i \wedge E^{\pm\pm} =0 \qquad 
\Leftrightarrow \qquad 
J_8 \wedge E^i \equiv J_8 \wedge \Pi^{\underline{m}} u_{\underline{m}}^i =0 
 \qquad 
\end{equation}
whose image on the worldsheet coincides with 
Eq. \p{Ei=0} 
\footnote{The precise argument goes as follows: Take  the integral  of
Eq. \p{du(c)} with an arbitrary 
10-dimensional test function $f(X)$. 
The integral 
of the forms  
$\hat{E}^i \wedge \hat{E}^{++}$ 
and  $\hat{E}^i \wedge \hat{E}^{--}$ multiplied by {\sl arbitrary} functions
$f(\hat{X})$ vanishes  
$$
\int\limits^{}_{{\cal M}^{(1+1)}} \hat{E}^i \wedge \hat{E}^{++} 
f(\hat{X})=0, \qquad 
\int\limits^{}_{{\cal M}^{(1+1)}} \hat{E}^i \wedge \hat{E}^{--} 
f(\hat{X})=0. 
$$
From the arbitrariness of $f(\hat{X})$ 
then both 2-forms are identically zero on the world sheet
$\hat{E}^i \wedge \hat{E}^{++}=
\hat{E}^i \wedge \hat{E}^{--}=0 $. And from the 
independence of  the pull--backs $\hat{E}^{++}$,  $\hat{E}^{--}$ 
indeed \p{Ei=0} follows.}
\begin{equation}\plabel{hEi=0}
\hat{E}^i \equiv \hat{\Pi}^{\underline{m}}(\xi ) \hat{u}_{\underline{m}}^i
(\xi ) =0 ~. \qquad 
\end{equation}

Now it becomes clear why the basis  
$E^{\underline{a}}$ \p{Ea9} 
\begin{equation}\plabel{EundaD9}
E^{\underline{a}} = \left( E^{++}, E^{--}, E^{i}
\right) \qquad 
E^{\pm\pm}= 
\Pi^{\underline{m}}u_{\underline{m}}^{\pm\pm}, \qquad  
E^{i}= 
\Pi^{\underline{m}}u_{\underline{m}}^{i},  \qquad  
\end{equation}
whose pull-back on the string worldsheet coincides with 
\p{Eunda},  is particularly convenient for the study of the coupled system. 
The dual basis $\nabla_{\underline{a}}$ \p{der} is 
constructed with the auxiliary moving frame variables  
\p{harmvD9}, \p{harmvD9IIB} 
$$
\nabla_{\underline{a}} = 
\left( \nabla_{++}, \nabla_{--} , \nabla_{{i}} 
\right)
\equiv 
u_{\underline{a}}^{~\underline{m}}\nabla_{\underline{m}}
$$
\begin{equation}\plabel{derIIB}
\nabla_{++}= 
{1 \over 2} u^{\underline{m}--}\nabla_{\underline{m}}, \qquad  
\nabla_{--} =  
{1 \over 2} u^{\underline{m}++}\nabla_{\underline{m}}, 
\qquad 
\nabla_{{i}}= - u^{\underline{m}i}\nabla_{\underline{m}}, 
\end{equation}
$$
\nabla_{\underline{m}}= \Pi^{-1}{}^{~n}_{\underline{m}}\partial_n .
$$

The decomposition of any form on the basis \p{Ea9}, \p{derIIB}   looks like  
\begin{equation}\plabel{hdec1}
d\Theta^{\underline{\mu }I}= 
E^{\pm\pm} \nabla_{\pm\pm}\Theta^{\underline{\mu}I} 
+ E^i \nabla_{i} \Theta^{\underline{\mu}I}, 
\end{equation}
($E^{\pm\pm} \nabla_{\pm\pm}\equiv E^{++} \nabla_{++} + E^{--} \nabla_{--}$) 
or 
\begin{equation}\plabel{hdec2}
E^{+qI} \equiv 
d\Theta^{\underline{\mu}I} v^{~+}_{\underline{\mu}q} 
= E^{\pm\pm} E^{+qI}_{\pm\pm} 
+ E^i E^{+qI}_{i},  
\end{equation}
\begin{equation}\plabel{hdec3}
E^{-\dot{q}I} \equiv 
d\Theta^{\underline{\mu}I} v^{~-}_{\underline{\mu}\dot{q}} 
= E^{\pm\pm} E^{-\dot{q}I}_{\pm\pm} 
+ E^i  E^{-\dot{q}I}_{i} 
\end{equation}
(cf. \p{ddec}). 
Due to \p{hEi=0}, only  the terms proportional to 
$E^{++}, E^{--}$ survive in the pull--backs 
of \p{hdec1}--\p{hdec3}
 on the superstring worldsheet
\begin{equation}\plabel{hdec1IIB}
d\hat{\Theta}^{\underline{\mu I}} (\xi )= 
\hat{E}^{\pm\pm} \left( \nabla_{\pm\pm}\Theta^{\underline{\mu I}} 
\right)(x(\xi)), 
\end{equation}
\begin{equation}\plabel{hdec2IIB}
\hat{E}^{+qI} \equiv 
d\hat{\Theta}^{\underline{\mu}I} \hat{v}^{~+}_{\underline{\mu}q} 
= \hat{E}^{\pm\pm} \hat{E}^{+qI}_{\pm\pm} ,
\end{equation}
\begin{equation}\plabel{hdec3IIB}
\hat{E}^{-\dot{q}I} \equiv 
d\hat{\Theta}^{\underline{\mu}I} \hat{v}^{~-}_{\underline{\mu}\dot{q}} 
= \hat{E}^{\pm\pm} \hat{E}^{-\dot{q}I}_{\pm\pm}. 
\end{equation}


An alternative way to represent Eqs. \p{hdec1IIB}, \p{hdec2IIB}, \p{hdec3IIB} 
is provided by the use of the current density \p{J81}, \p{J8Pi} and the 
equivalent version \p{du(c)} of Eq. \p{hEi=0}

\begin{equation}\plabel{hdec1J}
J_8 \wedge d{\Theta}^{\underline{\mu I}}= 
J_8 \wedge {E}^{\pm\pm} \nabla_{\pm\pm} \Theta^{\underline{\mu I}} (x), 
\end{equation}
\begin{equation}\plabel{hdec2J}
J_8 \wedge {E}^{+qI} \equiv 
J_8 \wedge d{\Theta}^{\underline{\mu}I} {v}^{~+}_{\underline{\mu}q} 
= J_8 \wedge {E}^{\pm\pm} {E}^{+qI}_{\pm\pm},  
\end{equation}
\begin{equation}\plabel{hdec3J}
J_8 \wedge {E}^{-\dot{q}I} \equiv 
J_8 \wedge d{\Theta}^{\underline{\mu}I} {v}^{~-}_{\underline{\mu}\dot{q}} 
= J_8 \wedge {E}^{\pm\pm} {E}^{-\dot{q}I}_{\pm\pm}.  
\end{equation}

\bigskip 

On the other hand, one can solve Eq. \p{du(c)} with respect to 
the current density. To this end we have to 
change the basis $\Pi^{\underline{m}} 
~\rightarrow E^{\underline{a}}=\Pi^{\underline{a}} u_{\underline{m}}^{\underline{a}}$ 
(see \p{Ea9}, \p{harmvD9}) in the expression 
\p{J8Pi} (remember that $det(u)=1$ due to \p{harmvD9}). Then the 
solution of \p{du(c)} becomes 

\begin{equation}\plabel{J8Eort}
J_8 = { 1 \over det(\Pi_{{r}}^{~\underline{s}})} 
(E^\perp)^{\wedge 8} ~{1 \over 2} \int_{{\cal M}^{1+1}} 
 \hat{E}^{++} \wedge \hat{E}^{--} 
\d^{10} \left( x - \hat{x} (\xi )\right), 
\end{equation}
where 
\begin{equation}\plabel{Eperp8}
(E^\perp)^{\wedge 8} \equiv 
 { 1 \over 8!} \e^{i_1 \ldots i_8} 
E^{i_1 } 
\wedge \ldots \wedge E^{i_1 } 
\end{equation}
is the local volume element of the space orthogonal to the worldsheet. 
The current form \p{J8Eort} includes an invariant 
on-shell superstring current 
 \begin{equation}\plabel{j(x)}
J_8 = (E^\perp)^{\wedge 8} j(x) , \qquad j(x) = 
{ 1 \over 2 det(\Pi_{{r}}^{~\underline{s}})} 
\int_{{\cal M}^{1+1}} 
 \hat{E}^{++} \wedge \hat{E}^{--} 
\d^{10} \left( x - \hat{x} (\xi )\right)~.
\end{equation}
Note that it can be written with the use 
of the 
Lorentz harmonics only.

The supersymmetric covariant volume can be decomposed as well 
in terms of the orthogonal volume form 
\begin{equation}\plabel{Pi8E8}
(\Pi )^{\wedge 10} \equiv 
(E^\perp)^{\wedge 8}  \wedge 
 { 1 \over 2} E^{++} \wedge E^{--} . 
\end{equation}

\bigskip

\subsubsection{Equations for auxiliary fields of super--D9--brane}

Variation with respect to the D9-brane Lagrange multiplier 
$Q_8$ yields the  identification 
of the auxiliary antisymmetric tensor field $F$ with the generalized 
field strength ${\cal F}$ of the Abelian gauge field $A$
\begin{equation}\plabel{delQ8(c)}
{\cal F}_2 \equiv dA- B_2 = F_2 \equiv  {1 \over 2} \Pi^{\underline{m}}
\wedge \Pi^{\underline{n}} 
F_{\underline{n}\underline{m}}~. 
\end{equation}

On the other hand, from
the variation with respect to the auxiliary antisymmetric tensor field 
$F_{\underline{n}\underline{m}}$ 
one obtains the expression for the Lagrange multiplier $Q_8$ 
\begin{equation}\plabel{delF(c)}
Q_8 = \Pi^{\wedge 8}_{\underline{n}\underline{m}}
\sqrt{|\eta + F|} (\eta + F)^{-1~\underline{n}\underline{m}}
\equiv - 
{ 1 \over \sqrt{|\eta + F|} } ~
[(\eta + F)_{\underline{.}\underline{.}}
 \Pi^{\underline{.}} ]^{\wedge 8 ~\underline{n}\underline{m}}
F_{\underline{n}\underline{m}}, 
\end{equation}
where  $\Pi^{\wedge 8}_{\underline{n}\underline{m}}$ is defined by 
\p{dualf} and (in a suggestive notation)
\begin{equation}
\plabel{dualfn}
[(\eta + F)_{\underline{.}\underline{.}}
 \Pi^{\underline{.}} ]^{\wedge 8 ~\underline{n}\underline{m}}
= 
{1 \over 2 ^. 8!} \e^{\underline{m}\underline{n}\underline{m}_1\ldots 
\underline{m}_{8}} 
(\eta + F)_{\underline{m}_1 \underline{n}_1} 
\Pi^{\underline{n}_1} 
\wedge ... \wedge 
(\eta + F)_{\underline{m}_8 \underline{n}_8} \Pi^{\underline{n}_{8}}. 
\qquad 
\end{equation}

The second form of \p{delF(c)} indicates that in the linearized approximation
with respect to the gauge fields one obtains 
\begin{equation}\plabel{delF(c)l}
Q_8 = - \Pi^{\wedge 8}_{\underline{n}\underline{m}}
F^{\underline{n}\underline{m}} + {\cal O} (F^2) 
\equiv - {1 \over 2} * F_2 + {\cal O} (F^2),  
\end{equation}
where $*$ denotes the $D=10$ Hodge operation and 
${\cal O} (F^2) $ includes terms of second and higher orders in 
the field $F_{\underline{n}\underline{m}}$.

\bigskip

\subsection{Dynamical bosonic equations:
Supersymmetric Born-Infeld equations with the  source.}

The supersymmetric generalization 
of the Born-Infeld dynamical equations 
\begin{equation}\plabel{delA(c)}
dQ_8 + d{\cal L}_8^{WZ-D7} = - dJ_8 
\end{equation}
follows from  variation with respect to the gauge field. 
Here we have to take into account the expression for 
$Q_8$ \p{delF(c)l}, the identification 
of $F$ with the gauge field strength \p{delQ8(c)} as well as the expression for the D7-brane Wess-Zumino term 
\begin{equation}\plabel{WZD7}
d{\cal L}_8^{WZ-D7} = 
 e^{\cal{F}} \wedge  R \vert_{9} , 
\qquad R = \oplus _{n=0}^{5} R_{2n+1} 
= 2i d\Theta^{2\underline{\nu} }  \wedge  d\Theta^{1\underline{\mu} } \wedge 
\oplus _{n=0}^{4} \hat{\s}^{(2n+1)}_{\underline{\nu}\underline{\mu} }. \qquad 
\end{equation}

Let us stress that, in contrast to the free Born--Infeld 
equation \p{delA},  Eq. \p{delA(c)}
has a right hand side produced by  the endpoints of the 
fundamental superstring. 

\bigskip 

Variation of the action with respect to $X^{\underline{m} }$  
yields 
\begin{eqnarray}\plabel{dXD9}
& {} & J_8  \wedge M^i_2 ~ u_{\underline{m} }^{~i} + 
\nonumber \\ 
& {} & + 2i  J_8 \wedge \left({E}^{2+}_{~{q}}\wedge {E}^{2+}_{~{q}} 
 u_{\underline{m} }^{--} - {E}^{1-}_{~\dot{q}}\wedge {E}^{1-}_{~\dot{q}}
u_{\underline{m} }^{++}  
 \right) +  \nonumber \\
& {} & +  \Pi^{\wedge 8}_{\underline{n}\underline{m}}
\sqrt{|\eta + F|} (\eta + F)^{-1~\underline{n}\underline{l}} 
\s_{\underline{l}}{}_{\underline{\mu}\underline{\nu}} 
 \wedge  
\left( d\Theta^{2\underline{\mu} }- d\Theta^{1\underline{\rho}}
h_{\underline{\rho}}^{~\underline{\mu} }\right) 
 \wedge  
\left( d\Theta^{2\underline{\nu} }- d\Theta^{1\underline{\e}}
h_{\underline{\e}}^{~\underline{\nu} }\right) + 
\nonumber \\ 
& {} & + dJ_8  \wedge
\left( -{1 \over 2} E^{\pm\pm} 
u^{\mp\mp \underline{n}} F_{\underline{n}\underline{m}  } 
+{1 \over 2} E^{++}u_{\underline{m} }^{--}    
-{1 \over 2} E^{--} u_{\underline{m} }^{++}   
\right) =  0 .  
\end{eqnarray}

The first line of Eq. \p{dXD9} contains the lifting to the super--D9--brane 
worldvolume 
of the 2-form $\hat{M}_2^{i}$ \p{hM2def} 
which enters the l.h.s. of the 
free superstring bosonic equations \p{M2=0} 
\begin{equation}\plabel{M2}
{M}_2^{i}\equiv 
1/2 {E}^{--} \wedge {f}^{++i} - 1/2 {E}^{++} \wedge {f}^{--i} 
+  2i {E}^{1+}_{~{q}}\wedge \g^i_{q\dot{q}}{E}^{1-}_{~\dot{q}}
- 2i {E}^{2+}_{~{q}}\wedge \g^i_{q\dot{q}}{E}^{2-}_{~\dot{q}}. 
\end{equation} 
The fourth line of Eq.\p{dXD9} again is the new input from the boundary.  

The second and third lines of Eq.\p{dXD9} vanish identically 
on the surface of the free fermionic equations of the free D9-brane and 
of the free superstring,  respectively. 
These are the Noether identities reflecting the 
diffeomorphism invariance of the free D9-brane and the 
free superstring actions.
 Hence, it is natural to postpone the discussion of  Eq. 
\p{dXD9} and turn to the fermionic equations for the coupled system.

\bigskip 

\subsection{Fermionic field equations}

The variation with respect to $\Theta^2$ produces the 
fermionic equation 
\begin{equation}\plabel{dTh2(c)}
\Pi^{\wedge 9}_{\underline{m}}
\sqrt{|\eta + F|} (\eta + F)^{-1~\underline{m}\underline{n}} 
\s_{\underline{n}}{}_{\underline{\mu}\underline{\nu}} 
 \wedge  
\left( d\Theta^{2\underline{\nu} }- d\Theta^{1\underline{\rho}}
h_{\underline{\rho}}^{~\underline{\nu} }\right) = 
- 2 J_8 \wedge E^{--} \wedge 
d\Theta^{2\underline{\nu} } 
~v_{\underline{\nu}q }^{~+}~
v_{\underline{\mu}q }^{~+}, 
\end{equation}
with the 
 r.h.s. localized at the worldsheet and 
proportional to the l.h.s. of the fermionic equations of 
the free type $IIB$ superstring (cf. \p{Th2str} and remembering that  
$ d\Theta^{2\underline{\nu} } 
~v_{\underline{\nu}q }^{~+}~\equiv E^{2+}_{~q}
$). 

The remaining  fermionic variation  
$\d\Theta^1$ produces an equation which includes the form $J_8$ 
with 
support localized at the worldsheet only: 

\begin{equation}\plabel{dTh1(c)}
J_8 \wedge \left( E^{++} \wedge E^{1-}_{~\dot{q}} 
v_{\underline{\mu}\dot{q} }^{~-}
- 
E^{--} \wedge E^{2+}_{~q } h_{\underline{\mu}}^{~\underline{\nu}} 
v_{\underline{\nu}q }^{~+}
\right) = 0. 
\end{equation}

This equation is worth a special consideration. 
For clearness, let us write its image on the worldsheet 
\begin{equation}\plabel{dTh1(c)0}
\hat{E}^{++} \wedge \hat{E}^{1-}_{~\dot{q}} 
\hat{v}_{\underline{\mu}\dot{q} }^{~-}
- 
\hat{E}^{--} \wedge \hat{E}^{2+}_{~q } \hat{h}_{\underline{\mu}}^{~\underline{\nu}} 
\hat{v}_{\underline{\nu}q }^{~+}
 = 0. 
\end{equation}
Contracting with inverse harmonics $v^{-\underline{\mu}}_{q}, 
v^{+\underline{\mu}}_{\dot{q}}$ and  
using the 'multiplication table' of the harmonics 
(\p{spharm1}) 
we arrive at the 
following  covariant $8+8$ splitting representation 
 for the  $16$ equations \p{dTh1(c)0}:  
\begin{equation}\plabel{dTh1(c)1}
\hat{E}^{++} \wedge \hat{E}^{1-}_{~\dot{q}} 
=
\hat{E}^{--} \wedge \hat{E}^{2+}_{~q } 
\hat{h}^{++}_{\dot{q}q}
\end{equation}
\begin{equation}\plabel{dTh1(c)2}
\hat{E}^{--} \wedge \hat{E}^{2+}_{~p } 
\hat{h}_{qp} = 0. 
\end{equation}
Here 

\begin{equation}\plabel{h++}
\hat{h}^{++}_{\dot{q}q} \equiv 
\hat{v}^{+\underline{\mu}}_{\dot{q} }
\hat{h}_{\underline{\mu}}^{~\underline{\nu}} 
\hat{v}_{\underline{\nu}q }^{~+}, \qquad  
\end{equation}
\begin{equation}\plabel{h-+}
\hat{h}_{pq} \equiv 
\hat{v}^{-\underline{\mu}}_{{p} }
\hat{h}_{\underline{\mu}}^{~\underline{\nu}} 
\hat{v}_{\underline{\nu}q }^{~+} \qquad  
\end{equation}
are the covariant $8 \time 8$ blocks of the image $\hat{h}$ 
of the Lorentz group valued (and hence invertible!) spin-tensor 
field $h$  \p{hinSpin}-\p{Id2} 
 \begin{equation}
\plabel{h16} 
h^{~\underline{\a}}_{\underline{\b}} \equiv 
v^{~\underline{\nu}}_{\underline{\b}} h^{~\underline{\nu}}_{\underline{\mu}} 
v^{~\underline{\a}}_{\underline{\mu}} 
\equiv 
\left( \matrix{ 
h_{qp} & h^{--}_{q\dot{p}} \cr 
h^{++}_{\dot{q}p} & \tilde{h}_{\dot{q}\dot{p}} \cr }
\right).   
\end{equation} 

Note that the source localized on the worldsheet of 
the open brane, as in \p{dTh2(c)} is characteristic for the system 
including a space-time filling brane. 
For the structure of the fermionic equations in 
the general case we refer to \cite{BK99}.

\bigskip

\section{Phases of the coupled system}
\setcounter{equation}{0}

It is useful to start with the fermionic equations of motion 
\p{dTh2(c)}, \p{dTh1(c)1}, \p{dTh1(c)2}. 

First of all we have to note that  in the generic phase    
{\sl there are no true (complete) Noether identities for the 
$\kappa$--symmetry} in the equations for the coupled system, 
 as all the $32$ fermionic equations are independent.

\subsection{Generic phase describing  
decoupled system and 
appearance of other phases}

In the generic case we shall assume that 
the matrix  $h_{qp}(X(\xi)) = \hat{h}_{qp}(\xi)$ is invertible 
($det( \hat{h}_{qp}) \not= 0$, for the case  $det( \hat{h}_{qp})= 0$ see 
Section 7.3). Then 
Eq. \p{dTh1(c)2} implies  
$\hat{E}^{--} \wedge \hat{E}^{2+}_{~p } = 0$ and immediately results 
in the reduction of the Eq. \p{dTh1(c)1}:  
\begin{equation}\plabel{fgen}
det( \hat{h}_{qp}) \not= 0 \qquad \Rightarrow \qquad 
\cases{ \hat{E}^{++} \wedge \hat{E}^{1-}_{~\dot{q}} 
= 0 \cr 
\hat{E}^{--} \wedge \hat{E}^{2+}_{~q } = 0  \cr } 
\end{equation}

The equations \p{fgen} have the same form as the free superstring equations
of motion \p{Th1str}, \p{Th2str}. As a result, the r.h.s. of 
Eq. \p{dTh2(c)} vanishes 
$$
det( \hat{h}_{qp}) \not= 0 \qquad \Rightarrow \qquad 
$$
\begin{equation}\plabel{dTh2(c)g}
\Pi^{\wedge 9}_{\underline{m}}
\sqrt{|\eta + F|} (\eta + F)^{-1~\underline{m}\underline{n}} 
\s_{\underline{n}}{}_{\underline{\mu}\underline{\nu}} 
 \wedge  
\left( d\Theta^{2\underline{\nu} }- d\Theta^{1\underline{\rho}}
h_{\underline{\rho}}^{~\underline{\nu} }\right) = 0
\end{equation}
which  coincides with the fermionic equation for the free super--D9--brane 
\p{dTh2eq}. Then 
the 
third line in the equations of motion for $X^{\underline{m}}$ coordinate 
fields 
\p{dXD9} vanishes, as it does in the free super--D9--brane case \p{dX}. 
As the second line in Eq. \p{dXD9} is zero due to the 
equations  \p{fgen} 
(e.g. $\hat{E}^{2+}_{~{q}}\wedge \hat{E}^{2+}_{~{q}} =$  
$(\hat{E}^{\pm\pm}\hat{E}^{~2+}_{\pm\pm {q}})
\wedge (\hat{E}^{\mp\mp}\hat{E}^{~2+}_{\mp\mp {q}}) =$ 
$- 2 \hat{E}^{--} \wedge \hat{E}^{2+}_{~{q}} ~\hat{E}^{~2+}_{-- {q}}
= 0$), in the generic case \p{fgen} 
the equations of motion for $X$ field \p{dXD9} become 
$$
det( \hat{h}_{qp}) \not= 0 \qquad \Rightarrow \qquad 
$$
\begin{equation}\plabel{dXD9g}
J_8 \wedge M^i_2 ~ u_{\underline{m} }^{~i} + 
 dJ_8  \wedge
\left( -{1 \over 2} E^{\pm\pm} 
u^{\mp\mp \underline{n}} F_{\underline{n}\underline{m}  } 
+{1 \over 2} E^{++}u_{\underline{m} }^{--}    
-{1 \over 2} E^{--} u_{\underline{m} }^{++}   
\right) =  0 . 
\end{equation}
 
Contracting equation \p{dXD9g}
with appropriate  harmonics \p{harmvD9}, one can split 
it into three covariant equations  
\begin{equation}\plabel{dXD9gi}
J_8 \wedge {M}^i_2  =  {1 \over 2} dJ_8  \wedge 
{E}^{\pm\pm} {F}^{\mp\mp i} , 
\end{equation}
\begin{equation}\plabel{dXD9g++}
dJ_8  \wedge E^{++} \left( 1 - {1 \over 2} F^{++ ~--}\right) = 0,
\end{equation}
\begin{equation}\plabel{dXD9g--}
dJ_8  \wedge E^{--} \left( 1 - {1 \over 2} F^{++ ~--}\right) = 0,
\end{equation}
where 
\begin{equation}\plabel{Fcomp}
{F}^{\mp\mp i} \equiv 
 u^{\underline{m}\pm\pm}  u^{\underline{n}i} F_{\underline{m}\underline{n}}, 
\qquad 
{F}^{++~--} \equiv 
 u^{\underline{m}++}  u^{\underline{n}--} F_{\underline{m}\underline{n}} 
\end{equation}
are contractions of the antisymmetric tensor field (gauge field strength) 
with the harmonics \p{harmvD9}. 

The l.h.s. of the first equation \p{dXD9gi} has support 
on the string world volume ${\cal M}^{1+1}$, while its r.h.s and 
all the equations \p{dXD9g++}, \p{dXD9g--} have support on the boundary of the string worldsheet $\partial{\cal M}^{1+1}$ only. 

An important observation is that 
the requirement for the superstring to have a nontrivial boundary 
$\partial{\cal M}^{1+1}\not= 0$ {\sl implies a specific  restriction 
for the image of the gauge field strength on the boundary of the 
string worldsheet}
\begin{equation}\plabel{F++--=2}
\partial{\cal M}^{1+1}\not= 0 \qquad \Rightarrow \qquad 
\hat{F}^{++~--} \vert_{\partial{\cal M}^{1+1}}
\equiv 
 \hat{u}^{\underline{m}++}  
\hat{u}^{\underline{n}--} F_{\underline{m}\underline{n}} 
\vert_{\partial{\cal M}^{1+1}} 
= 2 .
\end{equation}

Eqs. \p{F++--=2} can be regarded as 'boundary conditions' for the 
super--D9--brane gauge fields on  1-dimensional defects provided by 
the endpoints of 
 the fundamental superstring. 
Such boundary conditions describe a phase of the 
coupled system where the open superstring interacts 
with the D9-brane gauge fields through its endpoints.

However, the most general phase, which implies no restrictions \p{F++--=2}
on the image of the gauge field, is characterized by 
equations $dJ_8  \wedge E^{--}=0, ~dJ_8  \wedge E^{++}=0$ and 
$dJ_8=0$. This  means the conservation of the superstring current 
and thus implies that the superstring is closed
 \begin{equation}\plabel{F++--n=2}
 \hat{F}^{++~--} \vert_{\partial{\cal M}^{1+1}}\not =2 \quad \Rightarrow \quad
dJ_8=0 \quad \Rightarrow \quad
\partial{\cal M}^{1+1}= 0. 
\end{equation} 
The equations decouple and become  the equations 
of the free D9--brane and the ones of the free 
closed type $IIB$ superstring. 

Hence to arrive at the equations of a nontrivially coupled 
system of super--D9--brane and open  fundamental  
superstring 
we have to consider phases related to 
special 'boundary conditions' for the 
gauge fields on the string worldvolume or its boundary. 
The weakest form of such boundary conditions 
are provided by  \p{F++--=2}.

Below we  will describe some interesting 
phases characterized  by the 
boundary conditions formulated on the 
whole superstring worldsheet,  
but before that  some comments 
on the issues of 
$\kappa$--symmetry and supersymmetry seem to be important.

\bigskip 

\subsection{Issues of $\kappa$--symmetry and supersymmetry} 

\subsubsection{On $\kappa$--symmetry}

If one considers the field variation 
of the form \p{kD9}, \p{dkA} for the free D9-brane $\kappa$-symmetry 
transformation, one finds that they describe a  gauge symmetry 
of the coupled system as well, if the 
parameter $\kappa$  is restricted 
by 'boundary conditions' on the two dimensional defect 
(superstring worldsheet) 
\begin{equation}\plabel{kappab}
\kappa^{\underline{\mu}}(x(\xi)) \equiv 
\hat{\k}^{\underline{\mu}} (\xi ) =0.
\end{equation}

Thus we have a counterpart  of the $\kappa$-symmetry inherent 
to the host brane 
(D9-brane) in the coupled system. As the defect (string worldsheet) 
is a subset of measure zero in 10-dimensional space (D9-brane world volume) 
we still can use this restricted $\kappa$--symmetry to remove half of 
the degrees of 
freedom of the fermionic fields all over the D9-brane worldvolume except for 
the defect. 

At the level of Noether identities this 'restricted' 
$\kappa$--symmetry is reflected by the fact that the half of the fermionic 
equations \p{dTh1(c)} has nonzero support 
on the worldsheet only. 

For a system of low-dimensional  
intersecting branes and open branes ending on branes, which 
does not include the 
super--D9--brane or other space--time filling brane 
we will encounter an analogous situation 
where the $\kappa$--symmetries related to 
both branes should hold outside the intersection.

\bigskip 

However, we should note that, in the generic case \p{fgen}, all the 
$32$ variations of the Grassmann coordinates result in nontrivial equations. 
Thus we have {\sl no true counterpart} of the free brane $\kappa$-symmetry. 
Let us recall that the latter 
results in the dependence of half of the fermionic equations of the free 
superbrane. It is usually identified with the part (one-half) of  
target space supersymmetry preserved by the BPS state describing the brane 
(e.g. as the solitonic solutions of the supergravity theory).

\bigskip

\subsubsection{Bosonic and fermionic degrees of freedom and supersymmetry 
of the decoupled phase}

As  the general phase of our coupled system \p{fgen}, \p{F++--n=2} 
describes the decoupled super--D9--brane and closed type $IIB$ superstring, 
 it must exhibit the complete $D=10$ type $IIB$ supersymmetry. 
Supersymmetry (in a system with dimension $d >1$) 
requires the coincidence of the numbers of bosonic and fermionic 
degrees of freedom.  
We find it  instructive to consider how such a coincidence 
can be verified starting from the action of the coupled system, and 
to compare the verification with the one for the case of free branes.

\bigskip

In the free super--D9--brane case the $32$ fermionic fields 
$\Th^{\underline{\mu}I}$ can be split into $16$ physical and $16$ unphysical 
(pure gauge) ones. For our choice of the sign of the 
Wess-Zumino term \p{LWZ}
they can be identified with  $\Th^{\underline{\mu}2}$ and 
$\Th^{\underline{\mu}1}$, respectively. 

Then one can consider the equations of motion \p{dTh2eq} 
as restrictions of the physical degrees of freedom 
(collected in  $\Th^{\underline{\mu}2}$), while the pure gauge degrees of 
freedom ( $\Th^{\underline{\mu}1}$) can be removed completely by 
$\kappa$--symmetry \p{kD9}, i.e. we can fix a gauge 
$\Th^{\underline{\mu}1}=0$ (see \cite{schw1}).

\bigskip

A similar situation appears 
when one considers the  free superstring model, where one can identify the 
physical degrees of freedom with the set of 
$
\hat{\Th}^{1-}_{~\dot{q}} = 
\hat{\Th}^{\underline{\mu}1} \hat{v}^{~-}_{\underline{\mu}\dot{q}}, 
\hat{\Th}^{2+}_{~{q}} = 
\hat{\Th}^{\underline{\mu}2} \hat{v}^{~+}_{\underline{\mu}{q}}  
$ 
\p{analb}, while the remaining components 
$
\hat{\Th}^{1+}_{~{q}} = 
\hat{\Th}^{\underline{\mu}1} \hat{v}^{~+}_{\underline{\mu}{q}}, 
~\hat{\Th}^{2-}_{~\dot{q}} = 
\hat{\Th}^{\underline{\mu}2} \hat{v}^{~-}_{\underline{\mu}\dot{q}}, 
$ 
are pure gauge degrees of freedom with respect to the 
$\kappa$-symmetry whose irreducible form is given 
by \p{kappastr} (see \p{kappag}).

To calculate the number of degrees of freedom we have to remember that 
\begin{itemize}
\item 
pure gauge degrees of freedom are removed from the consideration {\sl completely}, 
\item 
the solution of second order equations of motion (appearing as a rule for  
bosonic fields, i.e. 
$\hat{X}^i(\xi )= \hat{X}^i(\tau, \s ) $ 
\p{analb}) for $n$ physical variables 
(extracted e.g. by fixing all the gauges) 
is characterized by $2n$ independent functions, which can be regarded as 
initial data for coordinates ($\hat{X}^i(0, \s )$)
and momenta (or velocities $\partial_\tau \hat{X}^i( 0, \s ) $),
\item the general solution of the first order equations 
(appearing as a rule for fermions, e.g. 
$\hat{\Th}^{1-}_{\dot{q}}(\tau, \s ) $, 
$\hat{\Th}^{2+}_{{q}}(\tau, \s ) $) 
is characterized by 
only $n$ functions, which can be identified with  the initial data 
for coordinates ( $\hat{\Th}^{1-}_{\dot{q}}(0, \s ) $, 
$\hat{\Th}^{2+}_{{q}}(0, \s ) $) 
which are identical to their momenta in this case. 
\end{itemize}

In this sense it is usually stated that 
$n$ physical (non pure gauge) fields 
satisfying the second order equations of motion 
{\sl carry $n$ degrees of freedom} 
(e.g. for $\hat{X}^i(\xi )$ ~~$n=(D-2)=8$), while 
$n$ physical fields 
satisfying the first order equations of motion 
 {\sl carry $n/2$ degrees of freedom}
(e.g. for $\hat{\Th}^{1-}_{\dot{q}}(\tau, \s ) $, 
$\hat{\Th}^{2+}_{{q}}(\tau, \s )$  
~~$n/2 = 2(D-2)/2 = 16/2=8$). 
This provides us with the same value $8$   
for the number of bosonic and fermionic degrees of freedom 
for both the free super--D9--brane and the free type $II$ superstring  
($8_{B}+8_{F}$). 

\bigskip

  If one starts from the action of a  
 coupled system similar to \p{SD9+IIB}, the counting should be performed 
in a slightly different 
manner,  
 because, as it was discussed above, {\sl we have
no  true $\kappa$--symmetry} in the general case. 
We still count $8$ physical bosonic degrees 
of freedom 
related to the super--D9--brane gauge field 
$A_{\underline{m}}(x)$  living in the whole bulk (super--D9--brane 
worldvolume),  
and 
$8$ physical bosonic degrees of freedom 
living on the 'defect' (superstring worldsheet) 
related to the orthogonal oscillations of the string 
 $\hat{X}^i(\xi )$. 

The 32 fermionic coordinate fields  
 ${\Th}^{\underline{\mu}1}(x), {\Th}^{\underline{\mu}2}(x)$
are restricted here by two sets of 16 equations 
\p{dTh2(c)}, \p{dTh1(c)}, with one set \p{dTh2(c)} involving the fields 
in the bulk (and also the source term with support on the worldsheet) 
and the other \p{dTh1(c)} with  support on the worldsheet only. 

As the field theoretical degrees of freedom are related to 
the general solution of homogeneous equations 
(in the light of the correspondence with the initial data described above), 
the presence (or absence) of the source with local support in the right 
hand part of 
the coupled equations is inessential and  
we can, in analogy with the free D9-brane case, 
treat the first equation \p{dTh2(c)} as the restriction on 
$16$ physical fermionic fields (say ${\Th}^{\underline{\mu}2}(x)$) 
{\sl in the bulk}. 
As mentioned above, the coupled system 
has a D9-brane-like $\kappa$--symmetry with the parameter 
${\k}^{\underline{\mu}}(x)$
restricted by the requirement that it should  vanish on the defect 
$\hat{\k}^{\underline{\mu}}(\xi ) \equiv 
{\k}^{\underline{\mu}}\left(x(\xi)\right)=0$. 
Thus we can use this kappa symmetry 
to remove the rest of the 16 fermionic fields 
(say ${\Th}^{\underline{\mu}1}(x)$) 
all over the bulk except at the defect. 

Thus all over the bulk {\sl including} the defect we 
have 8 bosonic fields, which are the components of 
$A_{\underline{m}}(x)$ modulo gauge symmetries  and 
$8=16/2$ fermionic fields, which can be identified with 
the on-shell content of ${\Th}^{\underline{\mu}2}(x)$. 

On the defect we have in addition the 16 components 
$\hat{\Th}^{\underline{\mu}1}(\xi )$, which are  
restricted (in the general case) by 16 first order equations 
\p{fgen} (or \p{Th1str}, \p{Th2str}) 
 and, thus, carry $8$ degrees of freedom. 
This is the same number of degrees of freedom 
as the one of  the orthogonal bosonic oscillations of superstring 
$\hat{X}^i (\xi) = \hat{X}^{\underline{m}}(\xi ) \hat{u}^i_{\underline{m}} 
(\xi )$. 
This explains why  our approach to the coupled system allows  
to describe a decoupled supersymmetric 
phase. 

It should be remembered (see Section 3.4) that the presence of boundaries 
breaks at least half of the target space supersymmetry.

\subsection{Phases implying restrictions on the gauge fields}
 
As mentioned in Section 8.1, 
the open fundamental superstring can be described  only when 
some restrictions on the image of the gauge field are implied. 
The simplest restriction is given by Eq. \p{F++--=2}. 
But it is possible to consider the phases where 
\p{F++--=2} appears as a consequence of  stronger restrictions 
which hold on the whole defect (string worldvolume), but not only on its 
boundary. 

An interesting property of such phases is that there   
an interdependence of the fermionic equations 
of motion emerges.
Such a dependence can be regarded as an additional  'weak'  counterpart 
of the $\kappa$-symmetry of the free superbrane actions. 

\subsubsection{Phases with  
less than $8$  
dependent fermionic equations}

The 
dependence of fermionic equations arises naturally 
when the 
matrix $\hat{h}_{pq}$ is degenerate:  
\begin{equation}\plabel{fdeg}
det( \hat{h}_{qp}) = 0 \qquad \Leftrightarrow \qquad 
r_h \equiv rank( \hat{h}_{qp}) < 8 
\end{equation} 
Then $ \hat{h}_{qp}$ may be represented through  
a set of 
$8 \times r_h$ rectangular matrices $S_q^{~I}$ 
\begin{equation}\plabel{fdegdec}
\hat{h}_{qp} = (\pm )S_q^{~I} S_p^{~I}, 
 \qquad q,p=1,\ldots 8, \qquad I= 1,\ldots r_h, \qquad 
r_h < 8 , 
\end{equation}
and 
Eq. \p{dTh1(c)2} implies only $r_h<8$ nontrivial 
relations 
\begin{equation}\plabel{ffeq}
0< r_h < 8  \qquad \Leftrightarrow \qquad  
\hat{E}^{--} \wedge \hat{E}^{2+}_{~q } S_q^{~I} = 0, \qquad I= 1,\ldots r_h ,
\end{equation}
while the remaining 
$8 - r_h$
fermionic equations are dependent. 

The general solution of Eq. \p{ffeq} differs 
from the expression for the fermionic equations 
of the free superstring $\hat{E}^{--} \wedge \hat{E}^{2+}_{~q }=0$ 
by the presence of 
$(8 - r_h)$ arbitrary fermionic two-forms (actually functions, 
as on the 
worldsheet any two-form is proportional to the volume 
$\hat{E}^{++} \wedge 
\hat{E}^{--}$) 
\begin{equation}\plabel{ffeqs}
0< r_h < 8  \qquad 
\Leftrightarrow \qquad 
\hat{E}^{--} \wedge \hat{E}^{2+}_{~q } = 
R^{~\tilde{J}}_q \hat{\Psi}_2{}_+^{~\tilde{J}}
 \qquad I= 1,\ldots r_h
\end{equation}
where the $8 \times (8- r_h)$ matrix $R^{~\tilde{J}}_q$ is composed 
of $8-r_h$ $SO(8)$ '$s$--vectors' 
which 
complete the set of $r_h$ $SO(8)$ $s$--vectors $S^{~{J}}_q$  to the 
complete basis in the $8$ dimensional space, i.e.
\begin{equation}\plabel{RS}
R^{~\tilde{J}}_q S^{~I}_q = 0 . 
\end{equation}
On the other hand, due to Eqs. \p{ffeqs}, \p{fdegdec},  
the $R^{~\tilde{J}}_q$ 
are the 'null-vectors' of the matrix 
$h_{pq}$ 
\begin{equation}\plabel{hR=0}
\hat{h}_{qp} R^{~\tilde{J}}_p =0 
\end{equation}
Thus, they may be used to write down the explicit form of the $8-r_h$ 
dependent fermionic equations 
\begin{equation}\plabel{depfe} 
 rank( \hat{h}_{qp}) = r_h < 8 \qquad \Rightarrow \qquad 
\left(\hat{E}^{--} \wedge \hat{E}^{2+}_{~q }\right) 
\hat{h}_{qp} R^{~\tilde{J}}_p 
\equiv 0 .
\end{equation}

\subsubsection{Nonperturbative phase with 8 dependent fermionic equations. } 

The case with the maximal number $8$ of dependent fermionic equations 
 appears when 
the matrix $h_{qp}$ 
vanishes at the defect ($\hat{h}_{qp} = 0$). 
As the complete matrix $h_{\underline{\a}}^{~\underline{\b}}$ 
\p{h16}, \p{Id1}, \p{Id2}
is Lorentz group valued \p{hinSpin} and, hence, nondegenerate   
($det (
h_{\underline{\a}}^{~\underline{\b}})\not= 0$), this implies that 
both antidiagonal $8 \times 8$ blocks  $h^{--}_{q\dot{p}}$, 
$h^{++}_{\dot{q}p}$ are nondegenerate
\begin{equation}\plabel{fmax}
\hat{h}_{qp} = 0, \qquad \Rightarrow  \qquad 
det(\hat{h}^{--}_{q\dot{p}}) \not=0, \qquad 
det(\hat{h}^{++}_{\dot{q}p}) \not=0. 
\end{equation}

In this case the fermionic equations 
\p{dTh1(c)1} are satisfied identically and thus we arrive at the 
system of $16+8=24$ nontrivial fermionic equations. 
The dependence of Eq. \p{dTh1(c)1} for the gauge field 
subject to the 'boundary conditions' \p{fmax} (see \p{Id1}, \p{Id2}) 
can be regarded as a counterpart of 8 $\kappa$--symmetries. 
Thus it could be expected that  ground state 
solutions 
corresponding to the BPS states preserving $1/4$ (i.e. $8$) of the 
$32$ target space supersymmetries should appear just in this phase.

It is important that the phase \p{fmax} is {\sl nonperturbative} 
in the sense that it has no a weak gauge field limit. 
Indeed, in the 
limit $F_{\underline{m}\underline{n}}
\rightarrow 0$ the spin-tensor 
$h_{\underline{\nu}}^{~\underline{\mu}}$ \p{hinSpin}, \p{Id1}
should tend to unity 
$h_{\underline{\nu}}^{~\underline{\mu}}= 
\d_{\underline{\nu}}^{~\underline{\mu}} + {\cal O}(F)$. 
As 
$\hat{v}^{-\underline{\mu}}_{{p} }
\hat{v}_{\underline{\mu}q }^{~+}=\d_{pq}$ (see Appendix A), the same is true 
for the $SO(8)$ $s$--tensor $h_{pq}$: 
 $h_{pq}= 
\d_{pq} + {\cal O}(F)$. Thus the condition \p{fmax} cannot be obtained 
in the weak field limit. 
This  reflects the fact that nontrivial coupling of the 
gauge field with string endpoints is described by this phase.

Another way to justify the above statements is to use \p{Id1}, 
\p{Id2} with the triangle matrix \p{h16}
 \begin{equation}
\plabel{h16tr} 
\hat{h}^{~\underline{\a}}_{\underline{\b}} \equiv 
\hat{v}^{~\underline{\nu}}_{\underline{\b}} 
\hat{h}^{~\underline{\nu}}_{\underline{\mu}} 
\hat{h}^{~\underline{\a}}_{\underline{\mu}} 
\equiv 
\left( \matrix{ 
0 & \hat{h}^{--}_{q\dot{p}} \cr 
\hat{h}^{++}_{\dot{q}p} & \tilde{\hat{h}}_{\dot{q}\dot{p}} \cr }
\right)  
\end{equation} 
and the explicit $SO(1,1)\times SO(8)$ invariant representation for 
$\sigma$--matrices (see Eq. \p{gammarep} in the Appendix A) to find that 
$\hat{h}_{pq} = 0$ implies 
(see Appendix C) 
\begin{equation}\plabel{F++--=2!}
\hat{F}^{++~--} 
\equiv 
 \hat{u}^{\underline{m}++}  
\hat{u}^{\underline{n}--} \hat{F}_{\underline{m}\underline{n}}  
= 2. 
\end{equation}
 Thus we see again that there is no weak gauge field limit, 
as the image of at least one of the gauge field strength components 
onto the string worldsheet has a finite value in the phase \p{fmax}. 
On the other hand, Eq. \p{F++--=2!} demonstrates that the condition 
\p{F++--=2} holds on the boundary of the worldsheet. 
Thus one can expect that 
this phase provides a natural possibility to describe the 
nontrivial coupling of the {\sl open} fundamental superstring with 
the D-brane gauge field. As we will prove below analysing 
the field equations, this is indeed the case.

\bigskip 

In the  'nonperturbative phase' \p{fmax} 
one of the fermionic equations 
\p{dTh1(c)2} is satisfied identically and 
thus we have only one nontrivial fermionic equation 
\p{dTh1(c)1} on the string worldsheet.  
Using the consequences \p{hdec2IIB} of Eq. \p{hEi=0} 
the 2-form equation \p{dTh1(c)1} can be decomposed 
 as
\begin{equation}\plabel{dTh1(c)11}
\hat{E}^{++} \wedge \hat{E}^{--} (
\hat{E}^{1~-}_{--\dot{q}} + \hat{E}^{2~+}_{++q } 
\hat{h}^{++}_{\dot{q}q}) = 0 .
\end{equation}
We find that it contains eight 0-form fermionic equations 
\begin{equation}\plabel{dTh1(c)12}
\hat{E}^{1~-}_{--\dot{q}}= -  \hat{E}^{2~+}_{++{q} } 
\hat{h}^{++}_{\dot{q}q}.   
\end{equation}
Another version of Eq.  \p{dTh1(c)12} is 
\begin{equation}\plabel{dTh1(c)14}
J_8 ~~ \left({E}^{1~-}_{--\dot{q}} + {E}^{2~+}_{++{q} } 
{h}^{++}_{\dot{q}q}\right) =0. 
\end{equation}
In the linearized approximation with respect to all fields 
{\sl except for the gauge field strength} $F$ ($h^{++}_{\dot{q}q}= {\cal O} 
(F)$) the equation \p{dTh1(c)12} becomes 
\begin{equation}\plabel{dTh1(c)13}
\partial_{--} \hat{\Th}^{-1}_{\dot{q}}= -  
\partial_{++} \hat{\Th}^{+2}_{{q}}   
\hat{h}^{++}_{\dot{q}q}.  
\end{equation}

\bigskip

One should remember that the free superstring fermionic equations 
\p{Th1str}, \p{Th2str}
as well as the equations \p{dTh1(c)1}, \p{dTh1(c)2} in the generic phase 
\p{fgen} imply 
$
det(\hat{h}_{qp})\not= 0, ~~ \Rightarrow 
~~ \hat{E}^{1~-}_{--\dot{q}}=0, ~~ \hat{E}^{2~+}_{++{q} }=0,  
$
whose linearized limit is 
$
\partial_{--} \hat{\Th}^{-1}_{\dot{q}}=0, ~~ 
\partial_{++} \hat{\Th}^{+2}_{{q}}= 0
$
with chiral fields as a solution 
$
\hat{\Th}^{-1}_{\dot{q}}=\hat{\Th}^{-1}_{\dot{q}} (\xi^{(++)} ), ~~ 
\hat{\Th}^{+2}_{{q}}=\hat{\Th}^{+2}_{{q}} (\xi^{(--)} )
$. 

The rest of the fermionic equations \p{dTh2(c)}
\begin{equation}\plabel{dTh2(c)m}
\Pi^{\wedge 9}_{\underline{m}}
\sqrt{|\eta + F|} (\eta + F)^{-1~\underline{m}\underline{n}} 
\s_{\underline{n}}{}_{\underline{\mu}\underline{\nu}} 
 \wedge  
\left( d\Theta^{2\underline{\nu} }- d\Theta^{1\underline{\rho}}
h_{\underline{\rho}}^{~\underline{\nu} }\right) = 
- 2 J_8 \wedge E^{--} \wedge E^{2+}_{~{q}} v_{\underline{\mu}q}^{~+} 
\end{equation}
$$
\equiv - 2 J_8 \wedge E^{++} \wedge E^{1-}_{~\dot{q}} (h^{++})^{-1}_{q\dot{q}}
v_{\underline{\mu}q}^{~+}
$$
has a nontrivial source localized on the superstring worldsheet. It is just 
proportional to the 
expression which vanishes in the free superstring case and in the generic 
phase, but remains nonzero in the phase \p{fmax}.

In the present case 
the relations  
\p{F++--=2} hold (see Eq. \p{F++--=2!}). Using these relations, 
straightforward but 
tedious calculations demonstrate that the projections 
of Eqs. \p{dXD9} for $\d X^{\underline{m}}$ onto the harmonics 
$u_{\underline{m}}^{\pm\pm}$ vanish identically here 
(these are Noether identities for reparametrization symmetry on the 
superstring worldvolume), while 
the projection onto $u_{\underline{m}}^{~i}$ results in 
\begin{equation}\plabel{dXi}
J_8 \wedge {M}_2^{i} = - {1 \over 2} dJ_8  \wedge
(E^{++} F^{--i} + E^{--} F^{++ i}) , 
\end{equation}
where $ F^{++ i},  F^{--i}$ and 
${M}_2^{i}$ are defined in Eqs. \p{Fcomp} and \p{M2} respectively.   
Eq. \p{dXi} differs from the one of the free superstring 
by the nonvanishing r.h.s.,  which has  support {\sl on the boundary} 
of the string 
worldsheet and 
describes the interaction with super-D9-brane gauge fields. 

The Born-Infeld equations has the form 
\p{delA(c)} (with \p{delF(c)}, \p{delQ8(c)} taken into account) 
and contains a nonvanishing source term $-dJ_8$. 

Thus, as  expected, the phase \p{fmax} describes the open 
fundamental superstring interacting with the super-D9-brane. 
The ends of the superstring carry the charge  of the super-D9-brane 
gauge field and provide the source for the supersymmetric Born-Infeld 
equation. 
Note that the source of the fermionic equations is localized on the whole 
worldsheet. 
This property is specific for the system including a space--time filling 
brane.

\bigskip


\section{Conclusion and outlook}
\def\theequation{\thesection.\arabic{equation}}
\setcounter{equation}{0}

In this paper we present the derivation of a  
complete set of supersymmetric equations for 
a coupled system consisting of the super--D9--brane and the open 
fundamental 
superstring 
'ending' on the D9-brane. 
To this end we construct a current distribution form 
$J_8$ which allows to write the action functional of superstring and D9-brane 
in similar forms, i.e. as an integral of a 10-form over the 10-dimensional 
space, 
 after the Grassmann coordinates of the superstring are identified 
with the images of the Grassmann coordinate fields of the super--D9--brane. 
We prove supersymmetric invariance of $J_8$.

{ The proposed way to construct the 
action for the coupled system of superstring and  space--time filling brane
requires the use of the 
moving frame (Lorentz harmonic) actions \cite{BZ,bpstv,bsv}
for the superstring. The reason is that  
its Lagrangian form (in distinction to the ones of the 
standard action \cite{gsw}) can be regarded as  pull--backs of some 
D-dimensional differential 2--form and, thus, the moving frame 
actions for the free superstring can be written easily as an 
integral  
over a D--dimensional manifold by means of the current density $J_8$. 
Just the existence of the moving frame formulation may motivate the 
{\sl formal}  
 lifting of the Lagrangian forms of the standard actions 
to D dimensions and their use for the description of the interaction with 
space--time filling branes and/or supergravity (see \cite{bbs} for bosonic 
branes).    
}

We obtain a complete supersymmetric system of the equations of 
motion for the coupled system of superstring and 
super--D9--brane.  Different phases of the coupled system 
are found.  
One of them can be regarded as generic, but describes the decoupled 
system of the closed superstring  and the super-D9-brane, 
while one of the others corresponds to a 
'singular' and nonperturbative 'boundary  condition' for the gauge field 
on the worldsheet. It describes the coupled system of the {\sl open} 
superstring interacting with the D9-branes 
and implies an interdependence of the fermionic equations of motion which 
can be regarded as a weak counterpart of the (additional) $\kappa$--symmetry.

\bigskip

The method proposed in {\cite{BK99} and elaborated in} this paper may also 
be applied to the   
construction of the action for a coupled  system containing any number 
$N_{2}$
of fundamental superstrings 
and any number $N_{2k}$ 
of type $IIB$ 
super-Dp-branes ($p=2k-1$) interacting with the super-D9-brane. 
In  the action of such a coupled system
\begin{equation}\plabel{acDp0}
S = \int_{{\cal M}^{1+9}} \left( {\cal L}_{10} + 
\sum_{k=1}^{4}\sum_{r_{2k}=1}^{N_{p=2k}} 
\int_{{\cal M}^{1+9}} J^{(r_{2k})}_{10-2k} \wedge {\cal L}^{(r_{2k})}_{2k} 
\right) + 
\end{equation}
$$
+ \sum_{s=1}^{N_{2}} 
\int_{{\cal M}^{1+9}} 
J^{(s)}_8 \wedge \left( {\cal L}^{(s)}_2 + dA \right)  
$$
${\cal L}_{10}= {\cal{L}}^0_{10} +  {\cal{L}}^1_{10} +  
{\cal{L}}^{WZ}_{10}$ is the Lagrangian form of the 
super-D9-brane action \p{SLLL}--\p{L1}, \p{LWZ}. 
${\cal L}^{(s)}_2$ represents the Lagrangian form \p{acIIB} for the 
$s$-th fundamental superstring lifted to the 9-brane world volume 
as  in \p{SD9+IIB},   
$J_8^{(s)}$ is the local supersymmetric current density 
\p{J81}, \p{J8Pi} for the $s$-th fundamental superstring. 
The latter is constructed 
with the help of the induced map of the worldsheet into the 
10-dimensional worldvolume of the super--D9--brane. Finally,   
${\cal L}^{(r)}_{2k}$ and $J^{(r)}_{10-2k}$ are the 
supersymmetric current density and a first order action functional 
for the $r$-th type $IIB$ super-Dp-brane with $p=2k-1=1,3,5,7$. 
The supersymmetric current density $J^{(r)}_{10-2k}$ 
$$
J^{(r)}_{10-2k} = (dx)^{\wedge 10-2k}_{{n}_1 \ldots  {n}_{2k}} 
J^{{n}_1  \ldots  {n}_{2k}}( x) =
$$
$$
 ={ 1 \over(10-2k)! (2k)!} \e_{{m}_1 \ldots {m}_{10-2k}  n_1 \ldots  {n}_{2k}}
  d{x}^{{m}_1 } \wedge \ldots \wedge
dx^{{m}_{10-2k}} \times 
$$
$$
\times \int_{{\cal M}^{1+2k}} 
d\hat{x}^{(r) {n}_1} (\zeta ) \wedge \ldots  \wedge 
d\hat{x}^{(r){n}_{2k}} (\xi ) 
\d^{10} \left(x - \hat{x}^{(r)} (\xi )\right)\equiv 
$$
\begin{equation}\plabel{J10-2k}
{} \end{equation}
$$
= { 1 \over (10-2k)! (2k)!} 
\e_{\underline{m}_1 \ldots \underline{m}_{10-2k} \underline{n}_1 \ldots 
\underline{n}_{2k} } { \Pi^{\underline{m}_1 } \wedge \ldots \wedge
\Pi^{\underline{m}_{10-2k} } 
\over det(\Pi_{{l}}^{~\underline{k}})} \times 
$$
$$
\times 
\int_{{\cal M}^{1+1}} 
\hat{\Pi}^{(r)\underline{n}_1}(\zeta ) \wedge \ldots \wedge
\hat{\Pi}^{(r)\underline{n}_{2k}} (\zeta )
\d^{10} \left( x - \hat{x}^{(r)} (\zeta)\right)
$$
is defined by  the 
induced map $x^m=\hat{x}^{(r)m} (\zeta)$ ($m=0,\ldots 9$) of the 
$r$-th Dp-brane worldvolume into the 10-dimensional worldvolume 
of the D9-brane, given by 
\begin{equation}\plabel{indmap}
\hat{X}^{(r)\underline{m}} (\zeta) = 
{X}^{\underline{m}}\left( \hat{x}^{(r)} (\zeta)\right)  \qquad 
\leftrightarrow \qquad 
\hat{x}^{(r)m} (\zeta)= 
x^m \left(\hat{X}^{(r)\underline{m}} (\zeta) 
\right). 
\end{equation}
The Lagrangian form ${\cal L}_{2k}$
of the first order action for the free super-Dp-brane with $p=2k$ can be 
found in 
\cite{bst}. Certainly the form of the interaction between branes, 
which can be introduced 
into ${\cal L}^{(r)}_{2k}$
by the boundary terms requires a separate consideration 
(e.g. one of the important points is the interaction 
with the D9-brane gauge field through the Wess-Zumino terms of Dp-branes).
We hope to return to these issues in a forthcoming  publication.

\bigskip 

It is worth mentioning that  the {\sl  super-D9-brane 
Lagrangian from ${\cal L}_{10}$ can be omitted form the action 
of the interacting system without loss of selfconsistency} 
{ (cf. \cite{BK99})}.
Thus one may obtain a supersymmetric description of the 
coupled  system  of fundamental superstrings and lower dimensional 
super-Dp-branes ($p=2k-1<9$), e.g. to the system of 
$N$ coincident super-D3-branes which is of interest 
for applications to gauge theory \cite{witten96,witten97},  
as well as  in the context of the Maldacena conjecture 
\cite{Maldacena}.

The only remaining trace of the D9-brane is  the 
existence of a map \p{indmap} of the super-Dp-brane ($p<9$)
worldvolume into a 10-dimensional space whose coordinates 
 are inert under a type $II$ supersymmetry. 
Thus the system contains 
an auxiliary all-enveloping 9-brane ({ '9-brane dominance'}). 
This means that we really do not need a space-time filling brane 
as a dynamical object and, thus, may be able to  extend our approach 
to the 
$D=10$ type $IIA$ and  $D=11$ cases, 
where such dynamical branes are not known. 

\bigskip

Another interesting direction for future study is 
to replace the action of the space-time filling brane by a counterpart of 
the group-manifold action for the corresponding supergravity theory 
(see \cite{grm}). Such an action also implies the map 
of a D-dimensional bosonic surface  into a space 
with $D$-bosonic dimensions, as the space time filling brane does. Thus 
we can define an induced map of the worldvolumes of superstrings and lower 
branes into the $D$--dimensional bosonic surface involved in  
the group manifold action and 
 construct the covariant action for the coupled system of 
intersecting superbranes {\sl and supergravity}.

In this respect the problem to construct  the counterpart 
of a group-manifold actions  for  
 the $D=10$ $type~II$ supergravity 
\cite{lst} and duality-symmetric $D=11$ supergravity \cite{bst}  
seems to be of  particular interest.

\bigskip

\section*{Acknowledgements} 

The authors are grateful to D. Sorokin and M. Tonin for interest in  
this paper and useful conversations and to 
R. Manvelyan, G. Mandal  for relevant discussions. 
One of the authors (I.B.) thanks the Austrian Science Foundation for the 
support within the project {\bf M472-TPH}. He acknowledges a 
partial support from the INTAS Grant {\bf 96-308} and the 
Ukrainian GKNT grant {\bf 2.5.1/52}.

 \newpage 

\section*{Appendix A. Properties of Lorentz harmonic variables}
\setcounter{equation}{0} 
\def\theequation{A.\arabic{equation}}

The  
Lorentz harmonic variables $u_{\underline{m}}^{~\underline{a}}$,  
$v_{\underline{\mu}}^{~\underline{\a}}$ 
parameterizing the coset 
$$
{SO(1,9) \over SO(1,1) \otimes SO(8)} 
$$
which are used in the geometric action like \p{acIIB} for 
$D=10$ superstring models. 
 
\subsection*{Vector   harmonics}

In any number of space--time dimensions the Lorentz harmonic variables
\cite{sok} which are appropriate 
 to adapt the target space vielbein to the string world volume
\cite{bpstv} 
are defined as $SO(1,D-1)$ group valued  $D \times D$ matrix 
\begin{equation}\plabel{harm}
u^{\underline{a}}_{\underline{m}} 
\equiv 
\left(u^{0}_{\underline{m}},
u^{i}_{\underline{m}}, 
u^{9}_{\underline{m}} \right) 
\equiv 
\left( {u^{++}_{\underline{m}} + u^{--}_{\underline{m}} \over 2},
u^{i}_{\underline{m}}, 
{u^{++}_{\underline{m}} - u^{--}_{\underline{m}} \over 2} \right) 
\qquad \in \qquad SO(1,D-1),  
\end{equation}
\begin{equation}\plabel{harmor}
\Leftrightarrow \qquad 
u^{\underline{a}}_{\underline{m}} 
u^{\underline{b}\underline{m}} =  
\eta^{\underline{a}\underline{b}} \equiv diag (+1,-1,...,-1).  
\end{equation}
In the light-like notations 
\begin{equation}\plabel{harmpmpm}
u^{0}_{\underline{m}}= {u^{++}_{\underline{m}} + u^{--}_{\underline{m}} 
\over 2}, \qquad 
u^{9}_{\underline{m}} = {u^{++}_{\underline{m}} - u^{--}_{\underline{m}} \over 2}  
\end{equation}
the flat Minkowski metric acquires the form 
\begin{equation}\plabel{metr}
\Leftrightarrow \qquad 
u^{\underline{a}}_{\underline{m}} 
u^{\underline{b}\underline{m}} =  
\eta^{\underline{a}\underline{b}} \equiv 
\left(\matrix{ 0 & 2  & 0 \cr
               2 & 0  & 0 \cr 
               0 & 0 & I_{8 \times 8} \cr}\right) , 
\end{equation}
and the orthogonality conditions look like \cite{sok}
$$ 
\Leftrightarrow \qquad 
u^{++}_{\underline{m}}
u^{++\underline{m}} = 0 , \qquad 
u^{--}_{\underline{m}} u^{--\underline{m}} =0, \qquad  
u^{++}_{\underline{m}} u^{--\underline{m}} =2 ,
$$
$$ 
u^{i}_{\underline{m}}
u^{++\underline{m}} = 0 , \qquad 
u^{i}_{\underline{m}}
u^{++\underline{m}} =0 , \qquad 
u^{i}_{\underline{m}}
u^{j\underline{m}} = - \d^{ij}. 
$$

\subsection*{A2. Spinor Lorentz harmonics}

For supersymmetric strings and branes 
we need to introduce the matrix  
$v_{\underline{\mu}}^{~\underline{\a}}$ 
which takes its values in the double covering 
$Spin(1,D-1)$ of the Lorentz group 
$SO(1,D-1)$ and provides the (minimal) spinor representation 
of the pseudo-rotation whose vector representation 
is given by the  
vector harmonics $u$ (spinor Lorentz harmonics \cite{B90,gds,BZ}). 
The latter fact implies the invariance of the 
gamma--matrices  with respect to the Lorentz group transformations 
described by $u$ and $v$ harmonics
\begin{equation}\plabel{harmsg}
v^{\underline{\a}}_{\underline{\mu}} \qquad \in \qquad Spin(1,D-1) 
\qquad \Leftrightarrow \qquad 
u^{\underline{a}}_{\underline{m}} 
\Gamma^{\underline{m}}_{\underline{\mu}\underline{\nu}} 
=  
v^{\underline{\a}}_{\underline{\mu}} 
\Gamma^{\underline{a}}_{\underline{\a}\underline{\b}} 
v^{\underline{\b}\underline{\nu}} , \qquad   
u^{\underline{a}}_{\underline{m}} 
\Gamma_{\underline{a}}^{\underline{\a}\underline{\b}} 
=  
v^{\underline{\a}}_{\underline{\mu}} 
\Gamma_{\underline{m}}^{\underline{\mu}\underline{\nu}} 
v^{\underline{\b}\underline{\nu}}.    
\end{equation}

\bigskip 

In this paper we use 
the $D=10$ spinor Lorentz harmonic variables 
$v_{\underline{\mu}}^{~\underline{\a}}$
parameterizing the coset 
$
Spin(1,9) /[Spin(1,1) \times SO(8)] 
$ 
\cite{BZ}, 
which are adequate for the description of $D=10$ superstrings. 
The splitting \p{harm} is reflected by the splitting of 
the $16\times 16$ Lorentz harmonic variables into two 
$16\times 8$ blocks 
$$
v^{~\underline{\a}}_{\underline{\mu}} 
\equiv 
( 
v^{~+ }_{\underline{\mu}q}, v^{~-}_{\underline{\mu}\dot{q}}) , 
\qquad \in \qquad Spin(1,9) 
$$
\begin{equation}\plabel{spharm}
v^{~\underline{\mu}}_{\underline{\a}} 
\equiv 
( v^{-\underline{\mu}}_{q}, v^{+\underline{\mu}}_{\dot{q}}) , 
\qquad \in \qquad Spin(1,9) 
\end{equation}
$$
v^{~\underline{\mu}}_{\underline{\a}} 
v^{~\underline{\b}}_{\underline{\mu}}  = 
\d_{\underline{\a}}^{~\underline{\b}}, \qquad  
v^{~\underline{\a}}_{\underline{\mu}}  
v^{~\underline{\nu}}_{\underline{\a}} = 
\d_{\underline{\mu}}^{~\underline{\nu}}, \qquad  \Leftrightarrow 
$$

\begin{equation}\plabel{spharm1}
v^{-\underline{\mu}}_{p}
v^{~+ }_{\underline{\mu}q} = \d_{pq}, 
\qquad 
v^{+\underline{\mu}}_{\dot{p}} v^{~-}_{\underline{\mu}\dot{q}} =
\d_{\dot{p}\dot{q}} ,  \qquad 
 v^{+\underline{\mu}}_{\dot{q}} v^{~+ }_{\underline{\mu}q} = 
0 =  v^{-\underline{\mu}}_{q} v^{~-}_{\underline{\mu}\dot{q}} ,
\qquad 
\end{equation}
$$
 \d_{\underline{\mu}}^{\underline{\nu}}
= v^{-\underline{\nu}}_{q} v^{~+}_{\underline{\mu}q} + 
 v^{+\underline{\nu}}_{\dot{q}} v^{~-}_{\underline{\mu}q}.
$$
To write in detail the relations \p{harmsg} 
 between spinor and vector 
Lorentz harmonics 
we need the explicit $SO(1,1) \times SO(8)$ invariant representation for the 
$D=10$ Majorana--Weyl gamma matrices $\sigma^{\underline{a}}$

$$
\s^{\underline{ 0}}_{\underline{\a}\underline{\b}}=
\hbox{ {\it diag}}(\delta _{{ qp}},
\delta_{{\dot q}{\dot p}})
= \tilde{\s }^{\underline{0}~\underline{\a}\underline{\b}} ,
\qquad \s^{\underline{9}}_{\underline{\a}\underline{\b}}=
\hbox{ {\it diag}} (\delta _{qp},
-\delta _{{\dot q}{\dot p}}) =
-\tilde{\s }^{\underline{9}~\underline{\a}\underline{\b}} ,
$$
\begin{equation}\plabel{gammarep}
\s^{i}_{\underline{\a}\underline{\b}} =
\left(\matrix{0 & \gamma ^{i}_{q\dot p}\cr
\tilde{\gamma}^{i}_{{\dot q} p} & 0\cr}
\right)
= - \tilde{\s }^{i~\underline{\a}\underline{\b}} ,
\end{equation}
$$
\s^{{++}}_{\underline{\a}\underline{\b}} \equiv
(\s^{\underline{ 0}}+
\s^{\underline{ 9}})_{\underline{\a}\underline{\b}}=
\hbox{ {\it diag}}(~2\delta _{qp},~ 0)
= -(\tilde{\s }^{\underline{ 0}}-
\tilde{\s }^{\underline{ 9}})^{\underline{\a}\underline{\b}} =
\tilde{\s }^{{--}~\underline{\a}\underline{\b}} ,
\qquad 
$$ 
$$
\s ^{{--}}_{\underline{\a}\underline{\b}}\equiv
(\s ^{\underline{ 0}}-\G ^{\underline{ 9}}
)_{\underline{\a}\underline{\b}}=
\hbox{ {\it diag}}(~0, ~2\delta
_{{\dot q}{\dot p}}) = (\tilde{\s }^{\underline{ 0}}+
\tilde{\s }^{\underline{ 9}})^{\underline{\a}\underline{\b}}
= \tilde{\s }^{{++}~\underline{\a}\underline{\b}}, 
$$
where $\g^i_{q\dot{q}} = \tilde{\g}^i_{\dot{q}q}$  are $8\times 8$
chiral gamma matrices
of the $SO(8)$ group.

Substituting \p{gammarep} we get from \p{harmsg}
\cite{gds,BZ,bpstv}
\begin{eqnarray}\plabel{harms10}
u^{{++}}_{\underline{m}} \s^{\underline{m}}_{\underline{\mu}\underline{\nu}}
=&
2 v^{~+}_{\underline{\mu}q} v^{~+}_{\underline{\mu}q} , \qquad 
u^{{++}}_{\underline{m}} 
\tilde{\s}^{\underline{m}~\underline{\mu}\underline{\nu}}=&
 2 v^{+\underline{\mu}}_{\dot q} v^{+\underline{\nu}}_{\dot q},
\qquad \nn
u^{{--}}_{\underline{m}} 
\s^{\underline{m}}_{\underline{\mu}\underline{\nu}}=&
2 v^{~-}_{\underline{\mu}\dot{q}} v^{~-}_{\underline{\mu}\dot{q}} , \qquad 
u^{{--}}_{\underline{m}} 
\tilde{\s}^{\underline{m}~\underline{\mu}\underline{\nu}}=&
 2 v^{-\underline{\mu}}_{q} v^{-\underline{\nu}}_{q},
\qquad \nn
u^{{i}}_{\underline{m}} 
\s^{\underline{m}}_{\underline{\mu}\underline{\nu}}=&
2 v^{~+}_{\{ \underline{\mu}q} \g^i_{q\dot{q}} 
v^{~-}_{\underline{\nu}\} \dot{q}} , \qquad 
u^{{i}}_{\underline{m}} 
\tilde{\s}^{\underline{m}~\underline{\mu}\underline{\nu}}=&
 - 2 v^{-\{ \underline{\mu}}_{q} \g^i_{q\dot{q}} 
v^{+\underline{\nu}\} }_{\dot{q}},
\qquad 
\end{eqnarray}
  $$
u^{{i}}_{\underline{m}}  \g^{i}_{q \dot q}  
= v^{+}_{q}
\tilde{\s}_{\underline{m}} v^{-}_{\dot q} =
- 
v^{-}_{q} \s_{\underline{m}} v^{+}_{\dot q},  \qquad
$$
$$
u^{{++}}_{\underline{m}}  \d_{pq} = v^{+}_q\tilde{\sigma}_{\underline{m}}
v^{+}_p, \qquad 
u^{{--}}_{\underline{m}}  \d_{\dot{p}\dot{q}} = 
v^{-}_{\dot{q}}\tilde{\sigma}_{\underline{m}}
v^{-}_{\dot{p}}. \qquad 
$$

The differentials of the harmonic variables are calculated 
easily by taking into account 
the conditions \p{harm}, \p{spharm}. 
For the vector harmonics this implies  
$$
 d u^{\underline{a}}_{\underline{m}} 
 u^{~\underline{b}\underline{m}} 
+
 u^{~\underline{a}\underline{m}} 
 d u^{\underline{b}}_{\underline{m}} 
= 0,
$$ 
whose solution is given by 
 \begin{equation}\plabel{hdif}
 d u^{~\underline{a}}_{\underline{m}} =
 u^{~\underline{b}}_{\underline{m}}
 \Om^{~\underline{a}}_{\underline{b}} (d)
 \qquad  \Leftrightarrow \qquad
 \cases {
 du^{~++}_{\underline{m}}
 = u^{~++}_{\underline{m}}  \om
 + u^{~i}_{\underline{m}} f^{++i} (d ) ,  \cr
 du^{~--}_{\underline{m}}
 = - u^{~--}_{\underline{m}}  \om
 + u^{~i}_{\underline{m}} f^{--i} (d ) , 
 \cr
 d u^{i}_{\underline{m}} = - u^{j}_{\underline{m}}  A^{ji} +
 {1\over 2} u_{\underline{m}}^{++} f^{--i} (d) +
 {1\over 2} u_{\underline{m}}^{--} f^{++i} (d) ,
\cr }
\end{equation}
where 
\begin{equation}\plabel{pC}
 \Om^{~\underline{a}}_{\underline{b}}
  \equiv
u_{\underline{b}}^{\underline{m}} d u^{\underline{a}}_{\underline{m}}
 = \pmatrix { 
\om &  0 & {1 \over \sqrt{2}} f^{--i} (d ) \cr
  0 & -\om & {1 \over \sqrt{2}} f^{++i} (d ) \cr
 {1 \over \sqrt{2}} f^{++i} (d ) &
 {1 \over \sqrt{2}} f^{--i} (d ) &
                         A^{ji} (d) \cr} , \qquad 
\Om^{\underline{a}\underline{b}}
\equiv 
\eta^{~\underline{a}\underline{c}}
\Om^{~\underline{b}}_{\underline{c}} =  
- \Om^{\underline{b}\underline{a}}
\end{equation}
are $SO(1,D-1)$ Cartan forms. They can be decomposed into the 
$SO(1,1)\times SO(8)$ covariant forms   
\begin{equation}\plabel{+i} 
f^{++i} \equiv
u^{++}_{\underline m} d u^{\underline{m}i} 
\end{equation}
\begin{equation}\plabel{-i}
 f^{--i} \equiv u^{--}_{\underline m} d u^{\underline{m}i} , 
\end{equation}
parameterizing the coset 
${SO(1,9) \over SO(1,1) \times SO(8)}$, 
the $SO(1,1)$ spin connection 

\begin{equation}\plabel{0}
\omega \equiv {1 \over 2} u^{--}_{\underline m} d
 u^{\underline  m~++}\; ,
\end{equation}
and $SO(8)$   connections  
(induced gauge fields)
\begin{equation}\plabel{ij}
 A^{ij} \equiv u^{i}_{\underline m} d u^{\underline m~j}\; .
\end{equation}

The Cartan forms \p{pC} 
satisfy the Maurer-Cartan equation
\begin{equation}\plabel{pMC}
d\Om^{\underline{a}~\underline{b}} - \Om^{\underline{a}}_{~\underline{c}}
\wedge \Om^{\underline{c}\underline{b}} = 0
\end{equation}
which appears as integrability condition for Eq.\p{hdif}. 
It 
has the form of a zero curvature condition. This reflects the fact 
 that the $SO(1,9)$ connections defined by the Cartan forms \p{pC} 
are trivial.  

The Maurer--Cartan equation \p{pMC} splits naturally into 
\begin{equation}\plabel{PC+} 
{\cal D} f^{++i}
\equiv d f^{++i} -  f^{++i} \wedge \omega + f^{++j} \wedge A^{ji} = 0  
\end{equation}
\begin{equation}\plabel{PC-}
 {\cal D} f^{--i} \equiv d  f^{--i} +  f^{--i} \wedge \omega +  f^{--j} \wedge A^{ji} = 0
\end{equation}
\begin{equation}\plabel{G}
 {\cal R} \equiv d \omega = {1\over 2} f^{--i} \wedge f^{++i}  
\end{equation}
\begin{equation}\plabel{R}
R^{ij}  \equiv d A^{ij} + A^{ik} \wedge A^{kj} = - f^{-[i} \wedge f^{+j]} 
\end{equation}
giving rise to the Peterson--Codazzi, Gauss and Ricci equations of  
classical Surface 
Theory (see \cite{Ei}).

The differentials of the spinor harmonics can be expressed in terms 
of the same Cartan forms \p{+i}--\p{ij}
\begin{equation}\plabel{vOmv}
dv_{\underline{\mu}}^{~\underline{\a}} 
= {1 \over 4} \Om^{\underline{a}\underline{b}} 
~v_{\underline{\mu}}^{~\underline{\b}} ~(\sigma_{\underline{a}\underline{b}})_{\underline{\b}}^{~\underline{\a}}.   
\end{equation} 
Using \p{gammarep} we can specify \p{vOmv} as
(cf. \cite{BZ}) 
\begin{equation}\plabel{0ij}
v^{-\underline{\mu}}_p d v^{~+}_{\underline{\mu}q} = 
{1\over 2} \d_{pq} \om - 
{1 \over 4} A^{ij} \g^{ij}{}_{pq} , 
\qquad 
v^{+\underline{\mu}}_{\dot{p}} d v^{~-}_{\underline{\mu}\dot{q}} = 
-{1\over 2} \d_{\dot{p}\dot{q}} \om - 
{1 \over 4} A^{ij} 
\tilde{\g}^{ij}{}_{\dot{p}\dot{q}}, \qquad 
\end{equation}
\begin{equation}\plabel{pmpmi1}
v^{+\underline{\mu}}_{\dot{p}} d v^{~+}_{\underline{\mu}q} = 
{1\over 2} f^{++i} \g^i_{q\dot{p}}, \qquad  
v^{-\underline{\mu}}_{q} d v^{~-}_{\underline{\mu}\dot{p}} = 
{1\over 2} f^{--i} \g^i_{q\dot{p}} 
\end{equation}
Note that in $D=10$ the relations between 
vector (v--), c--spinor and s--spinor representations of the $SO(8)$ 
connections 
have the completely symmetric form
$$ 
A_{pq} = 
{1 \over 4} A^{ij} \g^{ij}{}_{pq} ,   \qquad 
A_{\dot{p}\dot{q}} = {
1 \over 4} A^{ij} 
\tilde{\g}^{ij}{}_{\dot{p}\dot{q}}, \qquad 
$$
$$
A^{ij} = 
{1 \over 4} A_{pq} \g^{ij}{}_{pq} = 
{1 \over 4}  
A_{\dot{p}\dot{q}}
\tilde{\g}^{ij}{}_{\dot{p}\dot{q}}, \qquad 
$$
This expresses the well known triality property of the $SO(8)$ group 
(see e.g. \cite{gsw} and refs therein).

\bigskip 

\section*{Appendix B. Linearized bosonic equations of type $IIB$ superstring}
\setcounter{equation}{0} 
\def\theequation{B.\arabic{equation}}

Here we present the derivation of the linearized bosonic equations \p{linbeq} 
of the superstring from the  set of equations \p{Ei=0}, \p{M2=0}. 

In the gauge \p{kappag}, \p{staticg} fermionic inputs disappear from Eq. 
\p{M2=0}. Moreover, in the linearized approximation we can replace 
$\hat{E}^{\pm\pm}$ by the closed form $d\xi^{\pm\pm}$ 
(holonomic basis for the space tangent to the worldsheet) and solve 
the linearized Peterson-Codazzi equations 
\p{PC+}, \p{PC-}
\begin{equation}\plabel{PClin}
df^{++i}=0, \qquad df^{--i}=0  \qquad 
\end{equation}
in terms of two $SO(8)$--vector densities 
$k^{++i}, k^{--i}=0$ (infinitesimal parameters of the coset 
$SO(1,9)/[SO(1,1) \times SO(8)]$)
 \begin{equation}\plabel{PClins}
f^{++i}= 2dk^{++i}, \qquad f^{--i}= 2dk^{--i}. \qquad 
\end{equation}
Then the linearized form of the equations \p{Ei=0}, \p{M2=0} 
is 
 \begin{equation}\plabel{lEi=0}
dX^i - \xi^{++} dk^{--i} - \xi^{--i}dk^{++i}=0, \qquad 
\end{equation}
 \begin{equation}\plabel{lM2=0}
d\xi^{--i} \wedge dk^{++i} -  
d\xi^{++} \wedge dk^{--i} 
=0.
\qquad 
\end{equation}
Eq. \p{lM2=0} implies $$ 
\partial_{++}k^{++i}+ \partial_{--}k^{--i}=0, 
$$ 
 while the integrability conditions for Eq. \p{lEi=0} 
are 
 \begin{equation}\plabel{ilEi=0}
d\xi^{--i} \wedge dk^{++i} +   
d\xi^{++} \wedge dk^{--i} 
=0.
\qquad \rightarrow \qquad \partial_{++}k^{++i}- \partial_{--}k^{--i}=0
\end{equation}
Hence we have 
\begin{equation}\plabel{hi=0}
\partial_{++}k^{++i}=\partial_{--}k^{--i}=0.
\end{equation}

Now, extracting, e.g. the component of \p{lEi=0} proportional to 
$d\xi^{++}$ and taking into account \p{hi=0} one arrives at 
\begin{equation}\plabel{d++Xi}
\partial_{++}X^i = 
\xi^{++i} \partial_{++}k^{--i}.
\end{equation}
The $\partial_{--}$ derivative of Eq. \p{d++Xi}  again together with Eq. 
\p{hi=0} yields a relation  which
includes the $X^i$ field only 
\begin{equation}\plabel{d++--Xi}
\partial_{--}\partial_{++}
X^i = 
\xi^{++i} \partial_{++}\partial_{--}k^{--i}=0
\end{equation}
and is just the free equation \p{linbeq}.

\bigskip

\section*{Appendix C: 
The gauge field of D9-brane described by block--triangular 
spin-tensor $h$}
\setcounter{equation}{0} 
\def\theequation{C.\arabic{equation}}

Here we will present  the nontrivial solution of the 
characteristic equation \p{Id1}  
for the spin--tensor $h$ of the triangle form \p{h16tr}
 \begin{equation}
\plabel{h16tr1} 
\hat{h}^{~\underline{\a}}_{\underline{\b}} \equiv 
\hat{v}^{~\underline{\nu}}_{\underline{\b}} 
\hat{h}^{~\underline{\nu}}_{\underline{\mu}} 
\hat{h}^{~\underline{\a}}_{\underline{\mu}} 
\equiv 
\left( \matrix{ 
0 & \hat{h}^{--}_{q\dot{p}} \cr 
\hat{h}^{++}_{\dot{q}p} & \tilde{\hat{h}}_{\dot{q}\dot{p}} \cr }
\right)  \qquad \in \quad Spin(1,9). 
\end{equation}  

It corresponds to the $SO(1,9)$ valued matrix (cf. \p{Id2})
   \begin{equation}
\plabel{kab} 
k_{\underline{a}}^{~\underline{b}} 
\equiv 
u_{\underline{a}}^{~\underline{m}}  k_{\underline{m}}^{~\underline{n}}  
u_{\underline{n}}^{~\underline{~b}} 
\equiv \left( {k^{++}_{\underline{a}} + k^{--}_{\underline{a}} \over 2},
k^{i}_{\underline{a}}, 
{k^{++}_{\underline{a}} - k^{--}_{\underline{a}} \over 2} \right) 
\qquad \in \quad SO(1,9) 
 \end{equation}
 with the components 
   \begin{equation}
\plabel{k++} 
k_{\underline{a}}^{++} 
= {1 \over 2} 
\d_{\underline{a}}^{--} k^{++|++} , 
\qquad 
\end{equation} 
   \begin{equation}
\plabel{k--} 
k_{\underline{a}}^{--} 
= \d_{\underline{a}}^{++} 
{2 \over k^{++|++} } +
\d_{\underline{a}}^{--} 
{k^{++j} k^{++j} 
\over 2k^{++|++} } - 
\d_{\underline{a}}^{i} 
{2k^{ij} k^{++j} 
\over k^{++|++} } , 
\qquad 
\end{equation} 
   \begin{equation}
\plabel{ki} 
k_{\underline{a}}^{i} 
= 
{1 \over 2 } 
\d_{\underline{a}}^{--} 
k^{++i} - 
\d_{\underline{a}}^{j} 
k^{ji}. 
\qquad 
\end{equation} 
The matrix  $k^{ij}$, entering \p{k--}, \p{ki}, 
takes its values in the $SO(8)$ group: 
   \begin{equation}
\plabel{kkT} 
k^{ik}k^{jk} = \d^{ij} \qquad \Leftrightarrow \qquad k^{ij} \in SO(8). 
\qquad 
\end{equation} 

The nonvanishing $8\times 8$ blocks of the $16 \times 16$ matrix  $h$ 
\p{h16tr1} are related with the 
independent components 
$k^{++|++}$, $k^{++i}$, $k^{ij} \in SO(8)$ 
of the matrix \p{kab} by 
   \begin{equation}
\plabel{h--h--}
{h}^{--}_{q\dot{s}}
{h}^{--}_{p\dot{s}}= 
\d_{qp} 
{2 \over k^{++|++} } ,  
\qquad 
\end{equation} 
 \begin{equation}
\plabel{h--tih}
{h}^{--}_{q\dot{s}}
\tilde{h}_{\dot{q}\dot{s}}
= - 
\g^i_{q\dot{q}}
{k^{ij} k^{++j} 
\over 2k^{++|++} } , 
\qquad 
\end{equation} 
\begin{equation}
\plabel{h++h++}
{h}^{++}_{\dot{q}s}
{h}^{++}_{\dot{p}s}= 
\d_{\dot{q}\dot{p}} 
{k^{++|++} \over 2},  
\qquad 
\end{equation} 
  \begin{equation}
\plabel{tihtih}
\tilde{h}_{\dot{q}\dot{s}}
\tilde{h}_{\dot{p}\dot{s}}
= \d_{\dot{q}\dot{p}}
{k^{++j} k^{++j} 
\over 2k^{++|++} },  
\qquad 
\end{equation} 
\begin{equation}
\plabel{h++gih--}
{h}^{++}_{\dot{q}s}
\g^i_{s\dot{s}}
{h}^{--}_{q\dot{s}}=
- \g^j_{q\dot{q}} k^{ji}, 
\qquad 
\end{equation} 
   \begin{equation}
\plabel{h++gitih}
2{h}^{++}_{(\dot{q}|s}
\g^i_{s\dot{s}}
\tilde{h}_{|\dot{s})\dot{s}}=
\d_{\dot{q}\dot{p}} k^{++i}.  
\qquad 
\end{equation} 
These equations are produced by Eq. \p{Id1}
in the frame related to the harmonics \p{harmv}, \p{harms} 
of the fundamental  superstring.

The expression connecting the 
independent components  
$k^{++|++}$, $k^{++i}$, 
$k^{ij} \in SO(8)$ 
of the matrix  \p{kab}
with the components of   the antisymmetric tensor $F$ (which becomes 
the field strength of the
gauge field of the super--D9--brane on the  mass--shell)
$$
F_{\underline{a}\underline{b}} 
\equiv 
u_{\underline{a}}^a F_{ab} u_{\underline{~b}}^b = 
-F_{\underline{b}\underline{~a}} = (F^{--|++}, F^{++i}, F^{--i}, F^{ij}) 
$$ 
can be obtained from Eq. \p{Id2} in the frame related to the stringy harmonics
   \begin{equation}
\plabel{F--++}
F^{--|++} = 2, 
\qquad 
\end{equation} 
   \begin{equation}
\plabel{F++i}
F^{++i} = - {1\over 2} k^{++|++} F^{--i}, 
\qquad 
\end{equation} 
   \begin{equation}
\plabel{F--kF--}
F^{--j} 
k^{ji} = 
F^{--i} \Leftrightarrow 
F^{--j} 
(\d^{ji} -  
k^{ji}) = 0, 
\qquad 
\end{equation} 
   \begin{equation}
\plabel{F--k++=4}
F^{--i} 
k^{++i} = 4,
\qquad 
\end{equation} 
   \begin{equation}
\plabel{F--kk}
F^{--i} 
k^{++j} 
k^{++j} 
= 
- 4k^{ij} k^{++j} 
- 4F^{ij'} k^{j'j} k^{++j} 
\equiv  
- 4(\d^{ij} +F^{ij}) 4k^{jj'} k^{++j'} ,
\qquad 
\end{equation} 
   \begin{equation}
\plabel{Fij}
F^{ij'} 
(\d^{j'j}- k^{j'j}) 
= -  
(\d^{ij}+ k^{ij}) 
+ {1 \over 2}F^{--i} k^{++j}. 
\qquad 
\end{equation} 
 
In particular, the above results  demonstrate 
that Eq. $h_{pq}=0$ (\p{fmax} or \p{h16tr1}) indeed implies 
\p{F++--=2!} (see \p{F--++}).

\end{document}